\documentclass[twocolumn,groupedaddress,amsmath,aps,longbibliography,superscriptaddress]{revtex4-1}
\usepackage{amsmath,amsfonts}
\usepackage{dcolumn}
\usepackage{bm}
\usepackage{color}
\usepackage{multirow}
\usepackage{scrextend}
\usepackage{amssymb}
\usepackage{dsfont}

\usepackage{graphicx}
\usepackage{hyperref}
\usepackage{float} 
\usepackage[T1]{fontenc}
\usepackage[utf8]{inputenc}

\begin{document}


\title{Superconductivity in ternary Mg$_4$Pd$_7$As$_6$}



\author{Hanna Świątek}
\affiliation{Faculty of Applied Physics and Mathematics, 
Gdańsk University of Technology, Narutowicza 11/12, 80-233 Gdansk, Poland}
\affiliation{Advanced Materials Center, Gdańsk University of Technology, Narutowicza 11/12, 80-233 Gdansk, Poland}
\author{Sylwia Gutowska}
\email{sylwia.gutowska@univie.ac.at}
\affiliation{University of Vienna, Faculty of Physics, Computational Materials Physics, Kolingasse 14-16, 1090 Vienna, Austria}
\author{Michał J. Winiarski}
\affiliation{Faculty of Applied Physics and Mathematics, 
Gdańsk University of Technology, Narutowicza 11/12, 80-233 Gdansk, Poland}
\affiliation{Advanced Materials Center, Gdańsk University of Technology, Narutowicza 11/12, 80-233 Gdansk, Poland}
\author{Bartlomiej Wiendlocha}
\affiliation{AGH University of Krakow, Faculty of Physics and Applied Computer Science,
al. Mickiewicza 30, 30-059 Krakow, Poland}
\author{Tomasz Klimczuk}
\email{tomasz.klimczuk@pg.edu.pl}
\affiliation{Faculty of Applied Physics and Mathematics, 
Gdańsk University of Technology, Narutowicza 11/12, 80-233 Gdansk, Poland}
\affiliation{Advanced Materials Center, Gdańsk University of Technology, Narutowicza 11/12, 80-233 Gdansk, Poland}


\date{\today}

\begin{abstract}
We report the synthesis and characterization of a new compound Mg$_4$Pd$_7$As$_6$, which was found to be a superconductor with $T_c=5.45$~K. Powder X-ray diffraction confirms the U$_4$Re$_7$Si$_6$ structure (space group $Im$-$3m$, no. 229) with the lattice parameter $a$ = 8.2572(1)~\AA. Magnetization, specific heat, and electrical resistivity measurements indicate that it is a moderate-coupling ($\lambda = 0.72$) type-II superconductor. The electronic and phonon structures are calculated, highlighting the importance of antibonding Pd-As interactions in determining the properties of this material. The calculated electron-phonon copling parameter $\lambda = 0.76$ agrees very well with the experimental finding, which confirms the conventional pairing mechanism in Mg$_4$Pd$_7$As$_6$.
\end{abstract}


\maketitle

\section{Introduction}
Compounds formed from elements of the $s$ block, the $p$ block and the $d$ block of the periodic table offer the unique opportunity to study the hybridization between different orbitals and their influence on electronic structure, phonons, and superconductivity.
In recent years, the compounds of Pd, As and alkaline earth metal $A$ gained attention due to the series of compounds $A$Pd$_2$As$_2$ - the ending compounds of doped $A$Fe$_{2-x}$Pd$_x$As$_2$, which is an interesting case for studying the interplay of magnetism and superconductivity (with maximal $T_c=8.7$ K for $A$=Sr and $x=0.15$ \cite{apd2as2}). Furthermore, among the Pd-As compounds \cite{pd-as}, the hexagonal Pd$_2$As structure is superconducting below 1.7 K, while the cubic PdAs$_2$ does not show superconductivity. 
Among ternary compounds, we find the first Mg-based full Heusler compound MgPd$_2$Sb with a valence electron count VEC = 27, in which superconductivity has recently been reported with $T_c=2$ K \cite{mgpd2sb} and the non-superconducting half-Heusler compound MgPdSb \cite{mgpdsb}. 

Initially, motivated by this recent report on superconductivity, we set out to synthesize MgPd$_2$As. Preliminary synthesis attempts employing similar principles were unsuccessful and resulted in a multiphase sample, in which one of the phases was identified as an analog of Mg$_4$Ir$_7$As$_6$. 

The first report of a ternary intermetallic compound with a stoichiometry of 4:7:6 dates back to 1978, when Akselrud {\it et al.} \cite{akselrud1978-U4Re7Si6} reported the crystal structure of U$_4$Re$_7$Si$_6$. The U$_4$Re$_7$Si$_6$ compound crystallizes in the centrosymmetric body-centered cubic structure, the space group $Im$-$3m$ (no. 229), with two independent Re sites and the number of formula units per unit cell Z = 2. The structure is now a prototype of more than 70 intermetallic compounds and is quite unaffected by substitution, both in the case of different sizes and valence electron numbers of the atoms. Uranium atoms can be replaced by rare earths, magnesium, or group 4 transition metals, titanium, zirconium, and hafnium. In addition to rhenium, Tc, Co, Rh, Ir, Ru, and Os can be found at the T site. 

Most of the reported compounds are germanides and silicides, while compounds containing group 15 elements are much less common. To date, 16 antimonides, 5 arsenides, and 1 phosphide have been reported. The first arsenide, U$_4$Ru$_7$As$_6$, was reported by Noël {\it et al.} \cite{NOEL2000-U4Ru7As6} more than 20 years after the family was discovered. The first record of 4:7:6 ternary pnictides containing magnesium appeared in 2004 by Wurth {\it et al.} \cite{Wurth_Mg}. 
There, a crystal growth method was described, along with determination of the crystal structure, for one  arsenide, Yb$_4$Rh$_7$As$_6$, and three compounds with Mg: Mg$_4$Rh$_7$As$_6$, Mg$_4$Ir$_7$As$_6$ and Mg$_4$Rh$_7$P$_6$. The last compound is also the only phosphide reported so far.
Recently, Hirose {\it et al.} \cite{HIROSE-Yb4Ru7As6} reported the fifth arsenic-bearing compound, Yb$_4$Ru$_7$As$_6$, as the first ternary member of the Yb-Ru-As family. 
It was found to be a moderately heavy-fermion compound, exhibiting antiferromagnetic ordering below 2.5 K. In 2013, Schellenberg {\it et al.} \cite{Schellenberg-antimonides} reported a series of antimonides RE$_4$T$_7$Sb$_6$, where RE = Gd - Lu and T = Ru or Rh, including their structural characterization and magnetic properties of Ru-based members. The compounds containing Dy, Ho, Er or Tm exhibit Curie-Weiss paramagnetism with antiferromagnetic ordering occurring below 10 K, 5.1 K and 4 K for Dy$_4$Ru$_7$Sb$_6$, Ho$_4$Ru$_7$Sb$_6$ and Er$_4$Ru$_7$Sb$_6$, respectively. No ordering was observed for Tm$_4$Ru$_7$Sb$_6$. Yb$_4$Ru$_7$Sb$_6$ was found to be an intermediate valent compound based on its reduced magnetic moment value of 3.71 $\mu_B$/Yb. 
  
In this paper, we report the synthesis of Mg$_4$Pd$_7$As$_6$, crystal structure, experimental evidence of its superconductivity and finally analyze its electronic and phonon structures, calculated by the \textit{ab initio} methods, to explain the interplay between the unique atomic arrangement and the superconducting properties. To the best of our knowledge, Mg$_4$Pd$_7$As$_6$ is a new chemical compound and the first superconductor in the 4-7-6 family. 

\section{Methods}
The first intentional synthesis attempts carried out for Mg$_4$Pd$_7$As$_6$ using pure elements or employing a precursor Pd-As did not produce a single phase sample and the presence of large amounts of Pd$_2$As was detected in the pXRD patterns. For further attempts, Mg$_3$As$_2$ was first prepared as a precursor.
Magnesium metallic flakes (3N) and arsenic pieces (3N) were sealed together in an evacuated quartz tube in a ratio of approximately 3:2 with a slight excess of magnesium. The tube was heated to 580$^o$C and held there for 48 h. The obtained reddish material was ground into fine powder in a glovebox, because of its possible toxicity and instability in air. The precursor was then mixed with palladium powder (3N) and freshly ground arsenic in a ratio of 4:21:10, thoroughly ground and pressed into a pellet. After being placed in an alumina crucible it was sealed in an evacuated quartz tube, which was heated to 570$^o$C at a rate of 25$^o$C/h, held there for 50 h, and then air quenched to room temperature. The heating process was repeated for a slightly higher temperature of 590$^o$C to improve the crystallization of the sample. 
Powder x-ray diffraction (pXRD) measurements were performed using a Bruker D2 Phaser diffractometer with a LynxEye-XE detector, with Cu K$\alpha$ radiation ($\lambda=1.5406$~\AA). Structure refinement was carried out using GSAS-II software\cite{Toby:aj5212}.

The electronic structure was calculated using the density functional theory as implemented in Quantum Espresso (QE) \cite{qe,qe2}, with the use of ultrasoft pseudopotential. Pseudopotentials were generated with the help of QE, using input files from PSlibrary 0.3.1 \cite{pps}, with the exchange-correlation effects included within the Perdew-Burke-Ernzerhof generalized gradient approximation scheme (GGA)~\cite{pbe}. A mesh of $12^3$ $\bm k$ points in reciprocal space was used in the self-consistent cycle and $48^3$ points for calculations of the density of states. The wavefunctions and kinetic energy cut-offs were set to 80 Ry and 800 Ry, respectively.

The phonon structure was determined using the density functional perturbation theory \cite{baroni}. The mesh of $3^3$ $\bm q$ points was used, using an electronic structure calculated on the mesh of $6^3$ $\bm k$ points (it has been verified that the mesh of $8^3$ $\bm k$ points gives the same results at the $\Gamma$ point). On this basis and using the electronic eigenvectors calculated on the mesh of $12^3$ $\bm k$ points, the electron-phonon properties were calculated. The necessary integrals were calculated using the Gaussian method with a smearing parameter equal to 0.02 Ry.
 
 \section{Crystal structure}
 
\begin{figure}[b]
\includegraphics[width=0.5\textwidth]{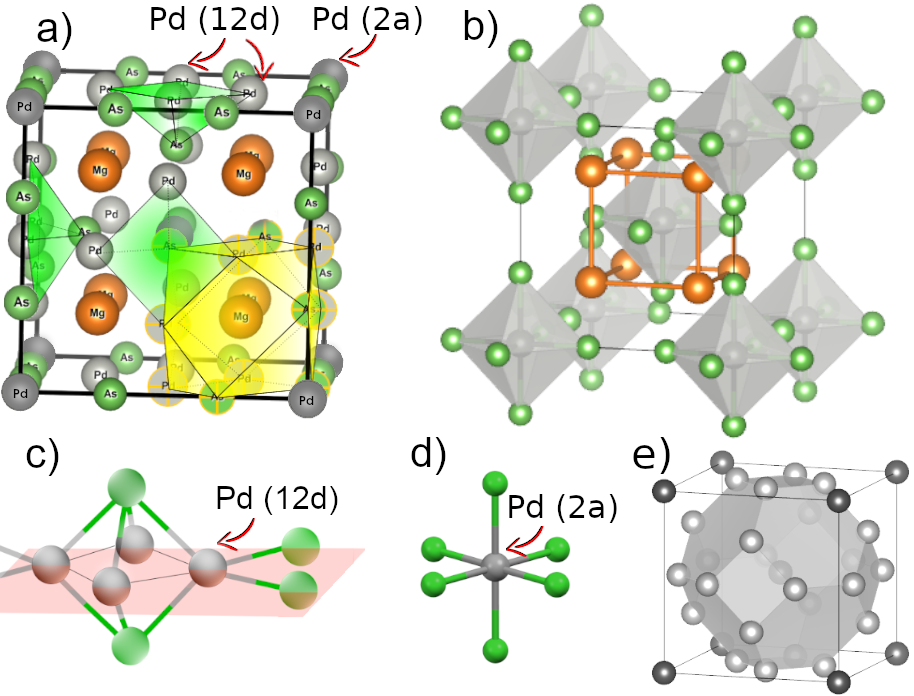} 
\caption{Crystal structure of Mg$_4$Pd$_7$As$_6$: a) the unit cell, were cuboctahedron of Pd(12d) and As atoms, centered on Mg atom is marked with yellow, while several AsPd(12d)$_4$ pyramids are marked with green, b) the unit cell with 
As$_6$ octahedra centered at Pd(2a) atom presented (for clarity the Pd(12d) atoms are removed), c)
Pd at \textit{12d} site and As schematic arrangement (bipyramids), d) Pd at \textit{2a} site and As arrangement (octahedra), e) unit cell with cube-truncated octahedron formed by  Pd(12d) presented (for clarity the atoms of Mg and As are removed).\label{fig:str}}
 \end{figure}

Mg$_4$Pd$_7$As$_6$ crystallizes in the body-centered U$_4$Re$_7$Si$_6$-type structure, space group $Im$-$3m$ No. 229, and its conventional unit cell is visualized in Fig.~\ref{fig:str}(a). It consists of two formula units: eight Mg atoms in Wyckoff position $(8c)$, twelve As atoms in (12e), and fourteen Pd atoms, 12 of which occupy position (12d), while the remaining 2 are in (2a) in the corners and in the center of the unit cell. The positions of Mg and Pd are fixed by symmetry, while the position of As is ($x,~0,~0$), where $x=0.31212$.

It is worth noting that if $x$ was equal to $0.31250$, then the As atom would be equally distanced from Pd(12d) and Pd(2a). On the other hand, if $x=0.25$, then all atoms would form squared atomic planes.

This crystal structure can be understood in multiple ways. Engel {\it et al.} \cite{engel} described it as a partially-filled deformed Cu$_3$Au-type structure, in which the Mg atoms are located on the gold sites, the As atoms are ordered on the copper sites, and the transition metal atoms are partially located on the copper sites, while also occupying the interstitial octahedral sites formed by the arsenic atoms.

In an alternative view, favored by later reports \cite{matar}, \cite{Leithe-structure} the structure can be described in terms of deformed cuboctahedra (marked with yellow in Fig.~\ref{fig:str}(a)) composed of six Pd(12d) and six As atoms, coordinating an Mg atom, and As octahedra (marked with gray in Fig.~\ref{fig:str}(b)) coordinating Pd(2a) atoms.

The nearest neighbors are As and Pd(12d), separated from each other by 2.6122\AA. Their bonding geometry is shown in Fig.~\ref{fig:str}(b). The coordination number of Pd(12d) is four because it is bonded to 4 As atoms.
In this case, a $sp^2d$ hybridization \cite{hybridization}  with the square planar atomic arrangement or sd$^3$ hybridization \cite{hybridization} with As atoms that form a tetrahedron around Pd is expected. 
Here, however, these tetrahedrons are strongly distorted -- the angle of As-Pd-As bonds is equal to 73$^o$ instead of 110$^o$ observed in a tetrahedron, and there are two As-Pd-As triangles perpendicular to each other. 
Alternatively, we can see this bonding as an As atom bound to four Pd (12d) atoms, forming bipyramids (centered at the (0.5, 0, 0) and (0.5, 0.5, 0) points of the conventional cell, shown in Fig.~\ref{fig:str}(a) with green), which can be seen as a squared distorted lattice. In this way the Pd(12d)$_6$-As$_6$ network is formed, schematically visualized in Fig.~\ref{fig:str}(c). When supplemented with a Pd (2a) atom and four Mg atoms placed in the space between pyramids, the Mg$_4$Pd$_7$As$_6$ structure is obtained. 

Atoms of As and Pd(2a) are distanced from each other by 2.6173 \AA. Their bonding geometry is marked with gray octahedra in Fig.~\ref{fig:str}(b) and schematically shown in Fig.~\ref{fig:str}(d).  The coordination number of Pd (2a) is six, it is bonded to six As atoms forming octahedral molecular geometry (such that there is an angle of 90$^o$ in the As-Pd-As triangles), characteristic of this coordination number and $sp^3d^2$ hybridization \cite{hybridization}. Mg are more distant from all other atoms and form a cube surrounding the central Pd (2a) atom, highlighting the link to the Heusler crystal structure.

Let us also highlight the other building blocks of this crystal structure. Figure \ref{fig:str}(e) presents the sublattice of Pd (12d) atoms, which form a cube-truncated octahedron, as was pointed out in the context of Yb$_4$Rh$_7$Ge$_6$ \cite{heying}. This feature, together with the Pd(2a)@As$_6$ octahedra, resembles those found in another ternary Mg-bearing compound, Mg$_{10}$Ir$_{19}$B$_{16}$, where they are composed of Ir(2) and Ir(1)-Mg atoms, respectively \cite{xu}.

\section{Experiment}
\begin{figure}[t]
\includegraphics[width=0.5\textwidth]{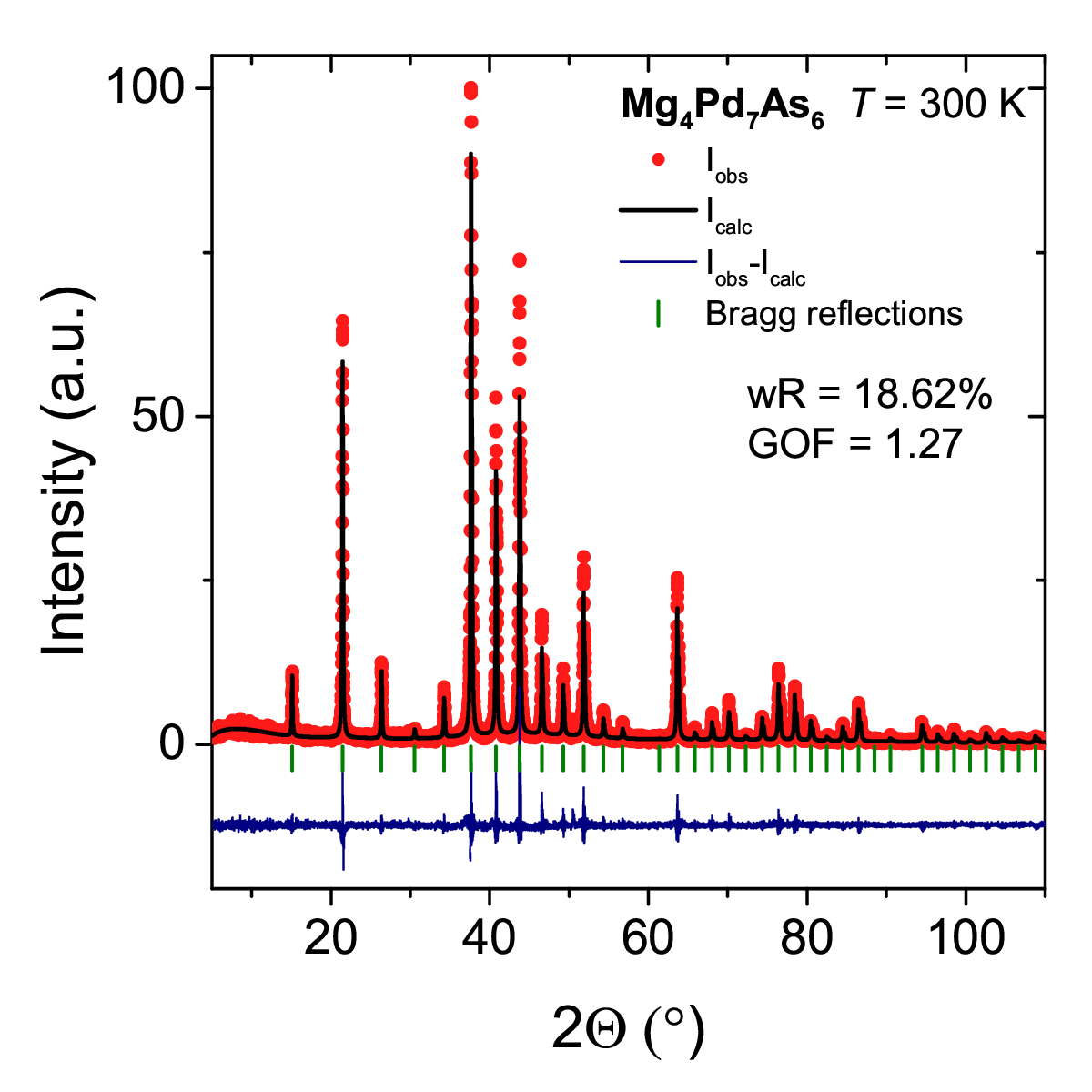} 
\caption{Powder x-ray diffraction pattern of Mg$_4$Pd$_7$As$_6$ (red points) with a Rietveld fit (black line). Green ticks mark the position of Bragg reflections of the \textit{Im-3m} Mg$_4$Pd$_7$As$_6$ phase. Blue line shows the difference between observed and calculated intensities.\label{fig:xrd}}
\end{figure}

Figure \ref{fig:xrd} shows the powder x-ray diffraction (pXRD) pattern of Mg$_4$Pd$_7$As$_6$ collected after the second step of synthesis. Rietveld analysis leads to the calculated lattice parameter of $a$ = 8.2570(1)~\AA, which, considering similar values of the atomic radius of the given transition metals, is considerably larger than for the analogue Mg$_4$T$_7$As$_6$ compounds, where $a$ is equal to 8.082(2) and 8.066(1)~\AA\  for Mg$_4$Ir$_7$As$_6$ and Mg$_4$Rh$_7$As$_6$, respectively \cite{mg4rh7as6-mg4ir7as6}. Details of the Rietveld refinement are presented in the Supplementary Material.

\begin{figure}
    \centering
    \includegraphics[width=0.5\textwidth]{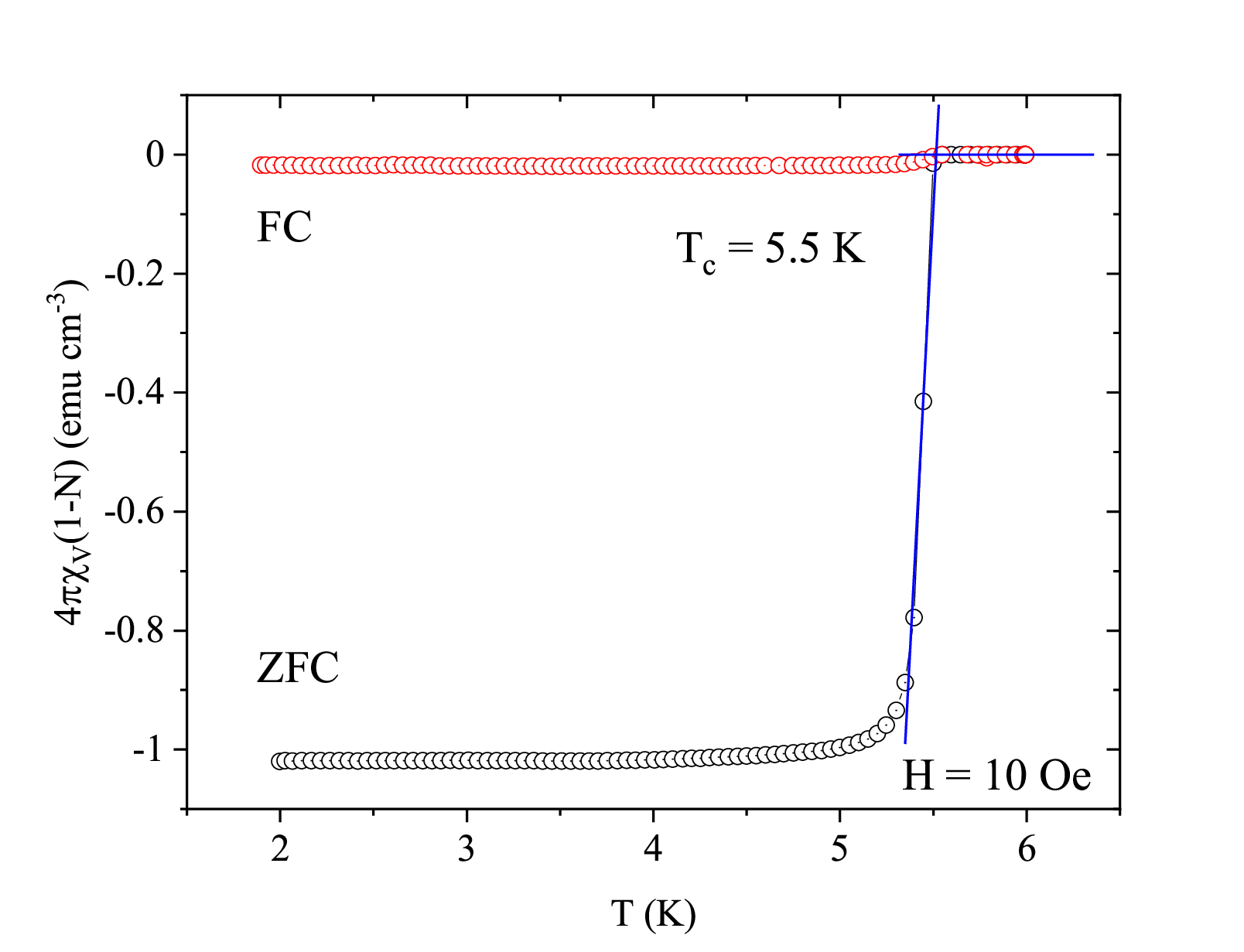} 

    \caption{Zero-field-cooled (ZFC) and field-cooled (FC) volume magnetic susceptibility of Mg$_4$Pd$_7$As$_6$ measured in a magnetic field of 10 Oe\label{fig:exp-m1}}
\end{figure}

The sample was characterized by measuring the dc magnetization in the low-temperature region in the zero-field-cooled (ZFC) and field-cooled (FC) modes. Measurements were carried out in the temperature range of 2-6 K in the applied field of 10 Oe. 
Figure \ref{fig:exp-m1} shows the volume magnetic susceptibility $\chi_v$. A strong diamagnetic signal is observed, indicating the superconducting character of the sample. The data were corrected for demagnetization effects, according to the $-4\pi\chi_\nu=\frac{1}{1-N}$ relation, with the demagnetization factor $N = 0.58$ calculated based on the subsequent isothermal $M(H)$ measurement at 1.9 K. 
The obtained value is consistent with those expected for the sample shape used in the measurement. The values of the corrected ZFC curve at the lowest temperatures reach -1, indicating full shielding. The significantly weaker FC signal indicates a low Meissner fraction and is expected for polycrystalline samples as a result of strong flux pinning at the grain boundaries. 
The transition temperature was determined as the intersection of the line lying along the steepest slope of the ZFC curve with the extrapolation of the normal state data and resulted in $T_c = 5.5$ K.

\begin{figure}
     \centering
     \includegraphics[width=0.5\textwidth]{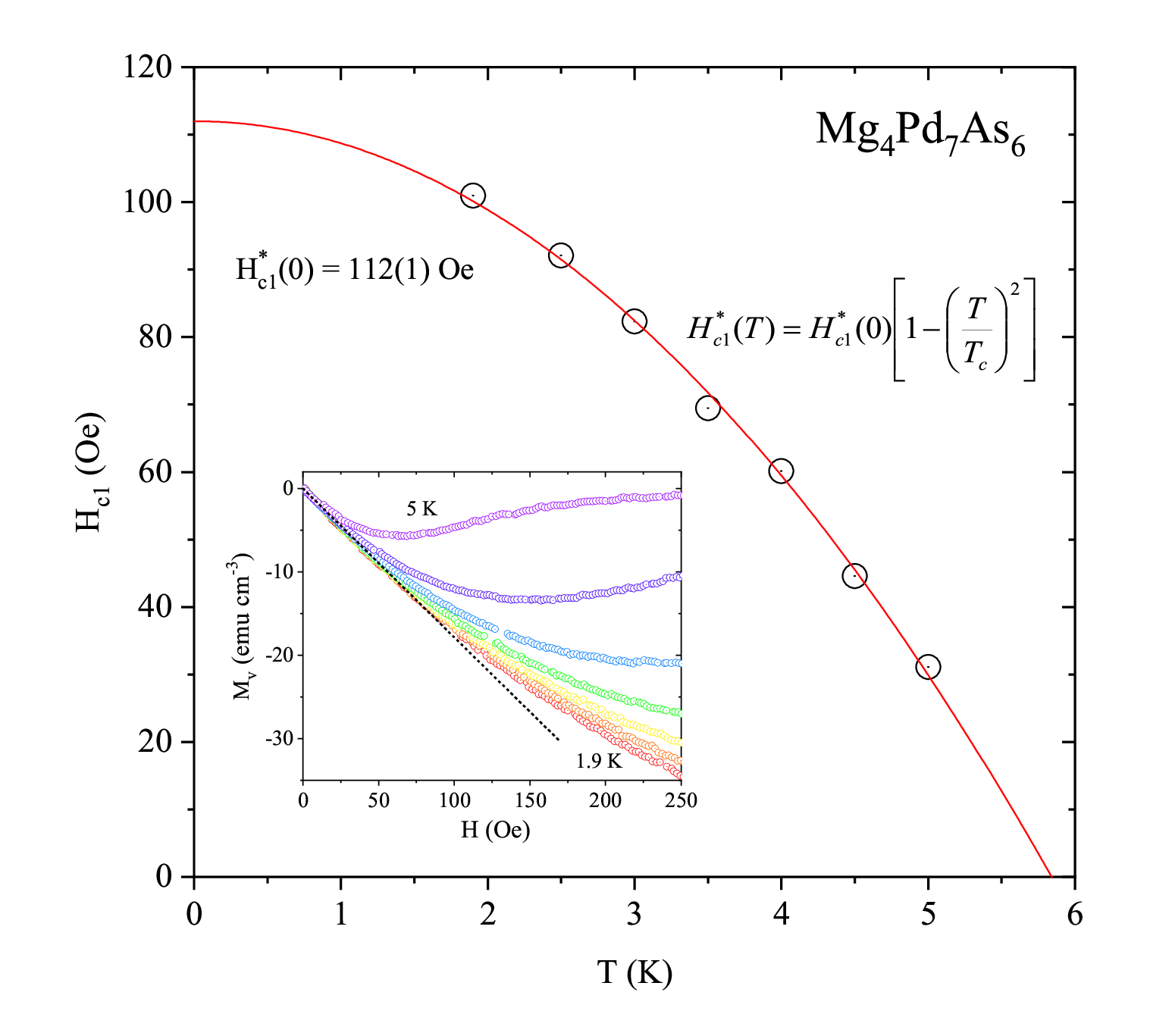}
     \caption{Temperature dependence of the lower critical field $H_{c1}$ determined from $M_{V}(H)$ measurements at various temperatures (shown in the inset)}
     \label{fig:exp-m2}
 \end{figure}

A series of low-field isothermal magnetization measurements was conducted for temperatures ranging from 1.9 to 5 K, to determine the lower critical field $H_{c1}(0)$. For each curve presented in the inset of Fig.~\ref{fig:exp-m2} the point of deviation from the initial linear response of the sample was determined. The obtained points are plotted in the main panel of Figure \ref{fig:exp-m2}, where they were fitted using the formula 
\begin{equation}
    H_{c1}^*(T)= H_{c1}^*(0)(1-\Big(\frac{T}{T_c}\Big)^2)
\end{equation}
The obtained value of $H_{c1}^*(0)$ was equal to 112(1) Oe, which after correcting for the demagnetization factor yielded $H_{c1}(0) = 267$ Oe. The shape of the magnetization curves (here, visible only for the higher temperatures) is typical of a type II superconductor.

\begin{figure}
     \centering
     \includegraphics[width=0.5\textwidth]{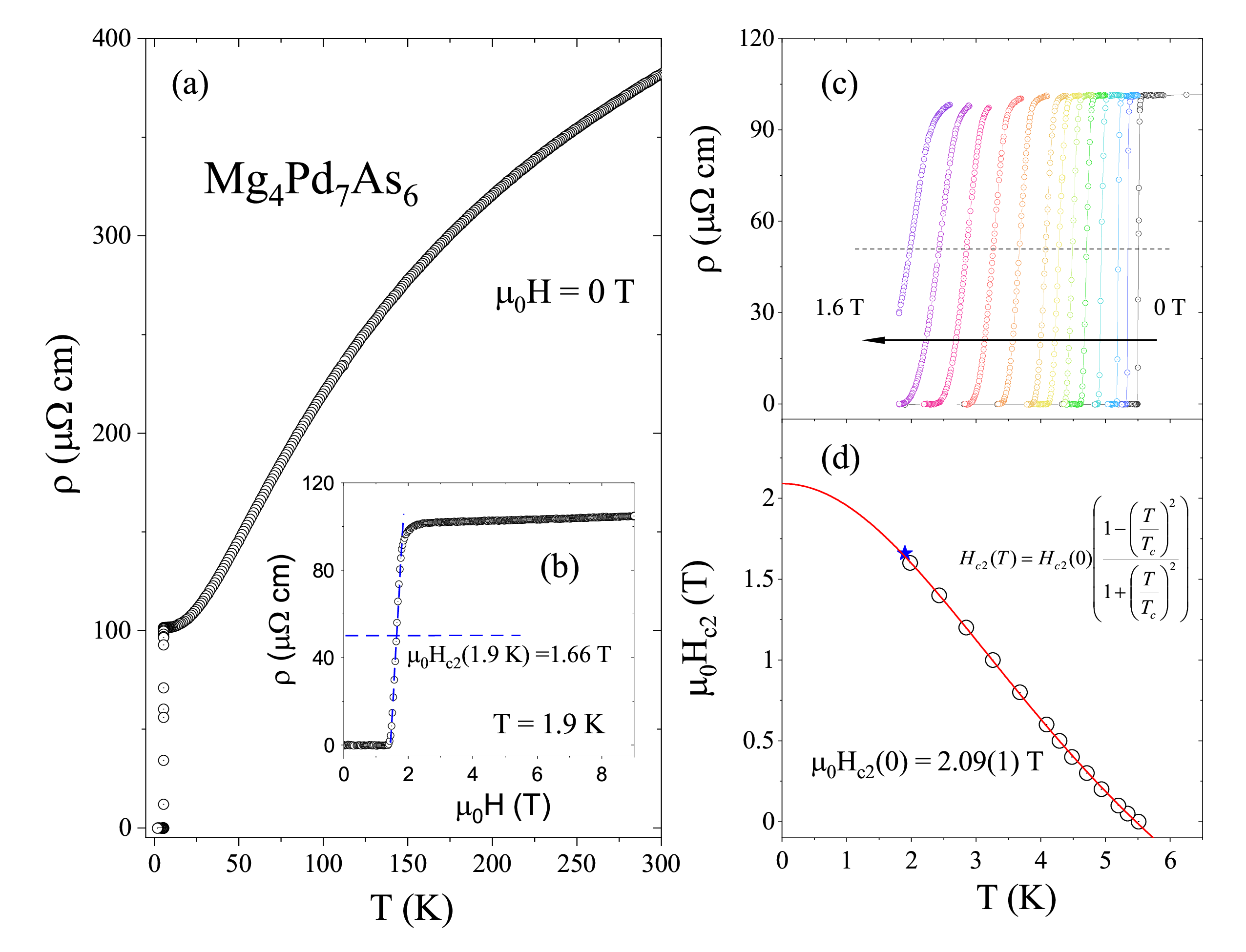}
     \caption{a) Temperature dependence of electrical resistivity of Mg$_4$Pd$_7$As$_6$ in the absence of external magnetic field; b) magnetic field dependence measured at a constant temperature of 1.9 K, with the estimated upper critical field marked by the intersection of the curve with a horizontal line at half-height; c) low temperature region of the electrical resistivity curves under applied magnetic fields between 0 and 1.6 T; d) temperature dependence of the upper critical field determined from resistivity measurements (marked with black circles) and the point estimated from $\rho(H)$ measurement (marked with a blue star).}
     \label{fig:exp-rho}
 \end{figure}

The temperature dependence of the electrical resistivity of the Mg$_4$Pd$_7$As$_6$ sample was measured in the range of 1.9 – 314 K, both with zero applied magnetic field and with a field of up to 1.6 T. The results of the zero-field measurement are presented in Fig.~\ref{fig:exp-rho}(a). In the full temperature range, the sample exhibits typical metallic behavior, with resistivity decreasing along with temperature ($\frac{d\rho}{dT} < 0$). The residual resistivity ratio of the sample is 4, which is typical for polycrystalline samples. The isothermal magnetic field dependence of resistivity, with data collected at 1.9 K, is presented in the inset. For this temperature, the upper critical field can be estimated to be 1.66 T.
Figure  \ref{fig:exp-rho}(c) shows the low temperature region, where a resistivity drop associated with a transition into a superconducting state can be observed. The critical temperature is determined as the mid-point of the resistivity drop. In the absence of an external magnetic field, the transition is sharp and $T_c = 5.51$ K, consistent with the magnetization measurement. 
For increasing magnetic field, the transition starts to broaden and shifts to lower temperatures, until superconductivity is fully suppressed at higher field values. The half-height temperature values of all curves were used to calculate the upper critical field $H_{c2}(0)$. The collected data points, presented in Fig.~\ref{fig:exp-rho}(d), were fitted using the Ginzburg-Landau function: 
\begin{equation}
    H_{c2}(T)= H_{c2}(0)\frac{1 - \Big(\frac{T}{T_{c}}\Big)^2}{1 +\Big(\frac{T}{T_{c}}\Big)^2}
\end{equation}
The obtained value was $\mu_0H_{c2}(0) = 2.09(1)$ T. The additional point estimated from the isothermal measurement at a temperature of 1.9 K is added to the plot and marked with a blue star. Its value is consistent with the fit line, and it falls marginally above it. 

From the lower and upper critical field values $\mu_0H_{c1}(0) = 26.7$ mT and $\mu_0H_{c2}(0) = 2.09$ T, we can estimate two other parameters important for characterizing superconductors, the coherence length $\xi_{GL}$ and the superconducting penetration depth $\lambda_{GL}$. From the Ginzburg-Landau formula $H_{c2} =\frac{\Phi_0}{2\pi\xi_{GL}^2}$, where $\Phi_0$ is the quantum flux, yields $\xi_{GL}(0) = 125$~\AA. Next, using the $H_{c1}=\frac{\Phi_0}{4\pi\lambda_{GL}^2}\ln(\frac{\lambda_{GL}}{\xi_{GL}})$ relation, we obtain 1174~\AA. Finally, the GL parameter $\kappa_{GL}=\frac{\lambda_{GL}}{\xi_{GL}}$ can be estimated. Inserting the obtained $\lambda_{GL}$ value, $\kappa_{GL}$ can be estimated to be $\kappa_{GL} = 9.4\gg \frac{1}{\sqrt{2}}$, which puts it firmly in the category of type-II superconductors. These parameters can then be used in the relation \(\ H_{c1}H_{c2} = ln(\kappa_{GL})H_{c}^2\) to calculate the thermodynamic critical field, $\mu_0H_{c}(0) = 50$ mT. 

 \begin{figure}[t]
     \centering
     \includegraphics[width=0.5\textwidth]{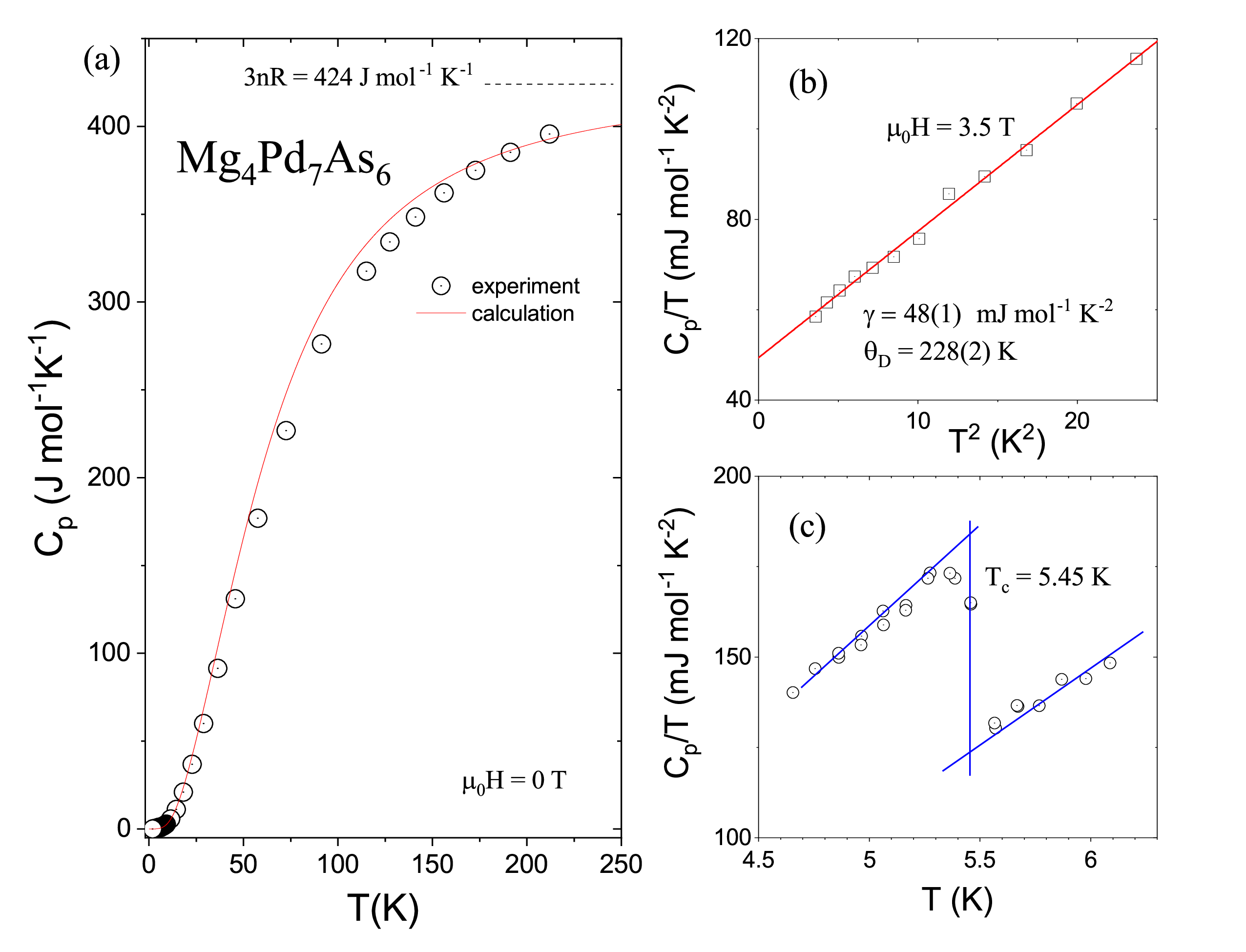}
     \caption{a) Zero-field specific heat in the full temperature range, where the red line shows the calculated $C_V$ values; b) normal-state $C_p/T$ vs $T^2$ in the low temperature region measured at $\mu_0H$ = 3.5 T. The red line corresponds to a linear fit used to determine electronic and phonon contributions to specific heat; c) the specific heat anomaly presented as $Cp/T$ vs $T$ in zero-field. The blue lines represent the equal-entropy construction used to determine the normalized specific heat jump.\label{fig:exp-cv1}}
 \end{figure}

To further characterize the superconducting properties of Mg$_4$Pd$_7$As$_6$, specific heat was measured in the temperature range of 1.9 – 210 K, with an applied magnetic field ranging from 0 to 3.5 T. Figure \ref{fig:exp-cv1}(a) presents the temperature dependence of the specific heat $C_p$ in a zero magnetic field (points). In the room temperature region, the values are expected to approach the Dulong-Petit limit of $3nR \simeq 424~\frac{\rm J}{{\rm mol} \cdot {\rm K}}$, where the number of atoms per formula unit is equal 17. 
The line in panel (a) shows the calculated constant-volume specific heat $C_V$, obtained using the formula~\cite{grimvall}
\begin{equation}
C_V = R\int_0^{\infty}F(\omega)\left(\frac{\hbar\omega}{k_BT}\right)^2\frac{\exp(\frac{\hbar\omega}{k_BT})}{(\exp(\frac{\hbar\omega}{k_BT})-1)^2.}
\end{equation}
where $F(\omega)$ is the theoretical  phonon density of states, discussed below. The agreement is good, validating the computed $F(\omega)$.

A closer look at the low temperature region is presented in Fig.~\ref{fig:exp-cv1}(b), with data collected under the applied field of $\mu_0H = 3.5$ T to analyze the normal-state data. Data are plotted as $C_p/T$ versus $T^2$ to demonstrate the expected linear dependence. The data was fitted in the range below 20 K$^2$ and the following parameters values were obtained: $\gamma = 48(1)~\frac{\rm mJ}{{\rm mol} \cdot {\rm K}^2}$ and $\beta = 2.80(6) \frac{\rm mJ}{{\rm mol} \cdot {\rm K}^4}$. Using the value of $\beta$, the Debye temperature was calculated to be $\theta_D = 228(2)$ K. This result is comparable to that of Yb$_4$Ru$_7$As$_6$, where $\theta_D = 250$ K \cite{HIROSE-Yb4Ru7As6} but lower than the isostructural compounds Yb$_4$T$_7$Ge$_6$ ($\theta_D \simeq 330$ K) and U$_4$Ru$_7$Ge$_6$ ($\theta_D = 330$ K) \cite{KATOH-Debye,Mentink-Debye}. 

\begin{table}[t]
    \caption{Superconducting parameters of Mg$_4$Pd$_7$As$_6$}
    \label{tab:supercond}
    \begin{ruledtabular}
    \begin{tabular}{c c c}
Parameter & Unit & Value\\ 
    \hline
$T_c$ & K & 5.45\\
$\mu_0H_{c1}(0)$ & mT & 26.7\\
$\mu_0H_{c2}(0)$ & T & 2.09\\
$\lambda$ & - & 0.72\\
$\xi_{GL}(0)$ & \AA & 125\\
$\lambda_{GL}(0)$ & \AA & 1174\\
$\kappa_{GL}$ & - & 9.4\\
$\gamma$ & $\frac{\rm mJ}{{\rm mol} \cdot {\rm K}^2}$ & 48\\
$\beta$ & $\frac{\rm mJ}{{\rm mol} \cdot {\rm K}^4}$ & 2.8\\
$\theta_D$ & K & 228\\
$\frac{\Delta C_p}{\gamma T_c}$ & - & 1.27\\
$RRR$ & - & 4
    \end{tabular}
\end{ruledtabular}
\end{table}

From the Debye temperature and $T_c$ we can further calculate the electron-phonon coupling parameter $\lambda$, using the inverted McMillan’s relation: 
\begin{equation}
\lambda=\frac{1.04+ \mu^*\ln(\frac{\theta_D}{1.45T_c})}{(1-0.62\mu^*)\ln(\frac{\theta_D}{1.45T_c})-1.04}
\end{equation}
where $\mu*$ is the Coulomb pseudopotential parameter, typically ranging between 0.10-0.15. Taking the commonly assumed value for conventional superconductors $\mu^* = 0.13$, $T_c = 5.45$ K and $\theta_D = 228$ K, the obtained value is $\lambda = 0.72$, indicating moderate coupling.

The specific heat jump at the transition temperature in the zero magnetic field is presented in Fig.~\ref{fig:exp-cv1}(c) as $C_p/T$ vs $T$. The characteristic lambda-shaped anomaly can be observed, signifying a transition into a superconducting state at a temperature of 5.45 K, marginally lower than the value obtained from the other measurements.

Finally, knowing the value of the Sommerfeld coefficient $\gamma$ and the transition temperature, the normalized superconducting jump $\frac{\Delta C_p}{ \gamma T_c}$ can be calculated using the equal-entropy construction (marked with blue lines in Fig. \ref{fig:exp-cv1}(c)). The obtained value is equal to 1.27, which is slightly lower than the 1.43 value predicted by the BCS theory for a weakly-coupled superconductor. It should be noted that $\frac{\Delta C_p}{ \gamma T_c}$ lower than 1.43 is commonly observed for many superconductors, i.e.  OsB$_{2}$ \cite{OsB2}.

At this point, an alternative way to calculate $\mu_0H_{c}(0)$ also becomes possible. Using the relation for the superconducting condensation energy \(\mu_0H_{c}^{2}(0)/2 = \iint(C(3.5T)-C(0T)/T dT\), which relates the difference between the data collected for the superconducting and normal state with the thermodynamic critical field, we obtain the value $\mu_0H_{c}(0)=59$ mT. The good agreement with the previously obtained value determined based on $\mu_0H_{c1}(0)$, $\mu_0H_{c2}(0)$ and $\kappa$ confirms the validity of all the derived parameters, collected in Table~\ref{tab:supercond}.

 \section{Electronic structure}
 
 \begin{figure}[t]
 \centering
\includegraphics[width=0.40\textwidth]{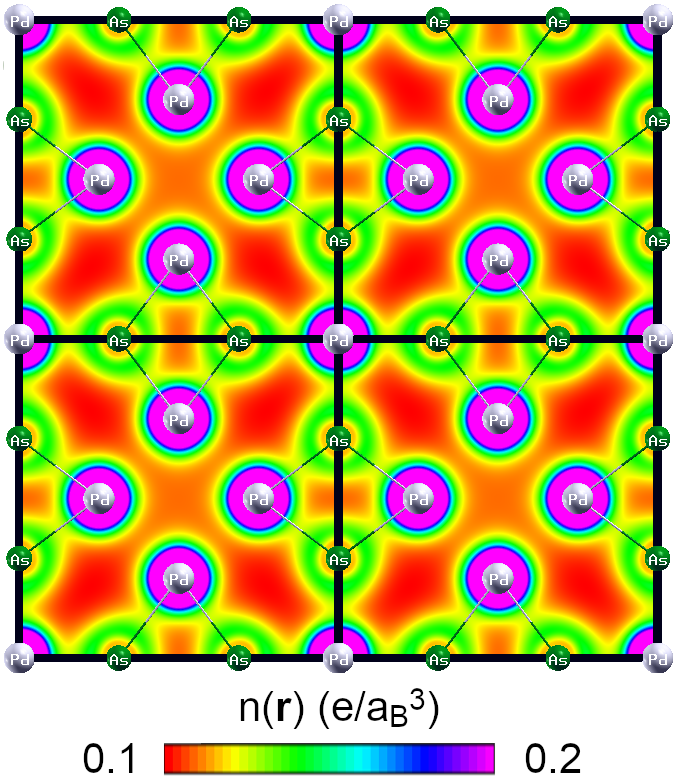} 
\caption{Electronic charge density of Mg$_4$Pd$_7$As$_6$ plotted in (100) plane.}\label{fig:charge}
 \end{figure}

 The calculated charge density on a face of the unit cell is shown in Fig.~\ref{fig:charge}.
 The atomic configurations of the compound components are as follows: Mg: [Ne]$3s^2$, Pd: [Kr]$4d^{10}$, As: [Ar]$3d^{10}4s^24p^3$.
 The Bader charge analysis made with the help of critic2 software \cite{critic} suggests that in the compound the valence electron occupation changes to: Mg: $3s^0$, Pd(12d): $4d^{10}5s^{0.69}$, Pd(2a): $4d^{10}5s^{0.22}$, As: $4s^24p^{3.57}$. 
 It agrees with the strong electropositive nature of Mg (its electronegativity on the Pauling scale is equal to 1.31), which is why this atom contributes its electrons to both Pd and As. The latter atoms have similar electronegativities (2.20 for Pd and 2.18 for As) and form bonds, visible in the charge density plot in Fig.~\ref{fig:charge}.
 
\begin{figure}[t]
\includegraphics[width=.5\textwidth]{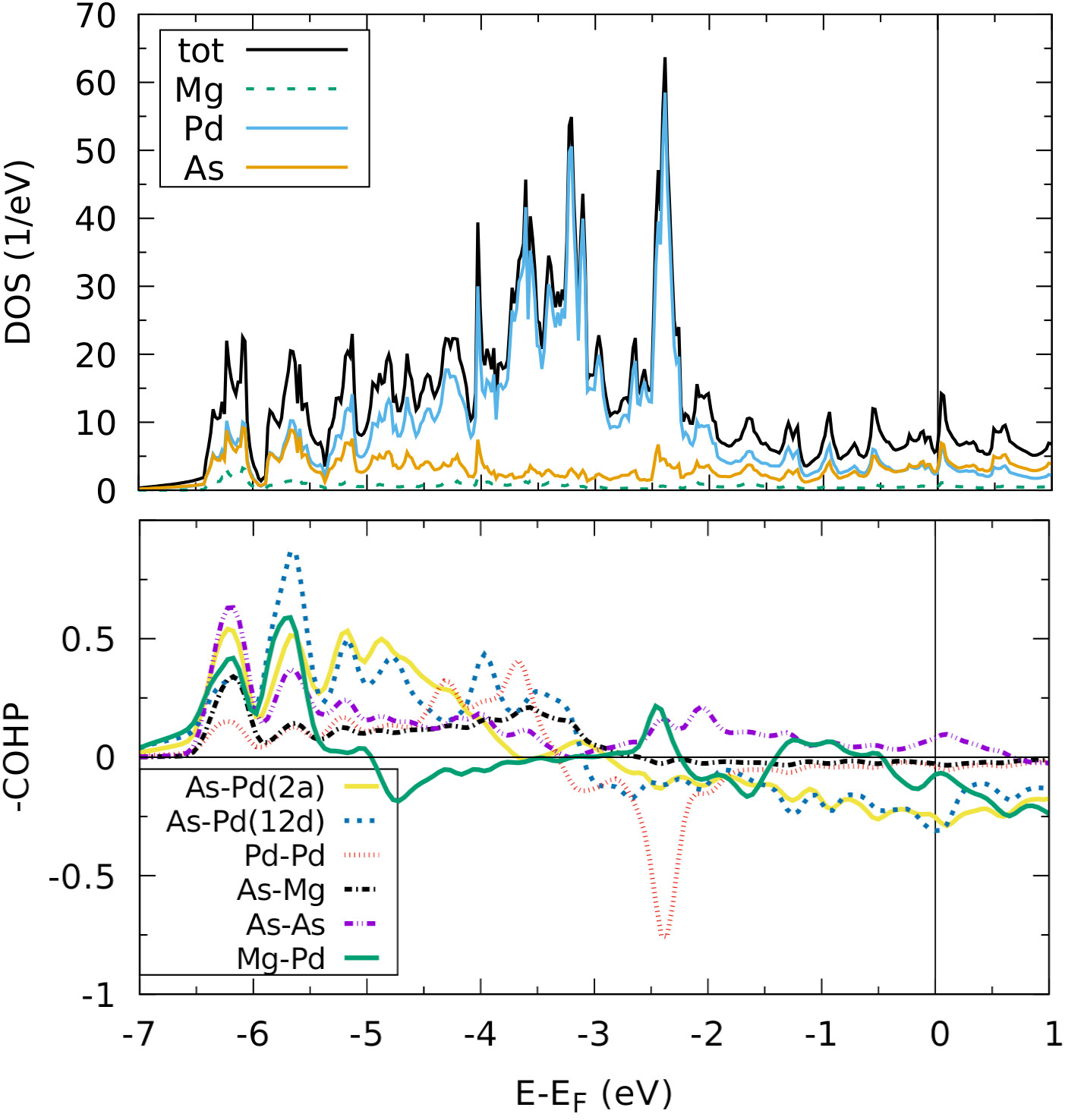} 
\caption{a) Total and partial electronic density of states of Mg$_4$Pd$_7$As$_6$; b) its crystal orbital Hamilton populations.\label{fig:dos}}
 \end{figure}

The electronic density of states of Mg$_4$Pd$_7$As$_6$ is shown in Fig.~\ref{fig:dos}. It is dominated mainly by the $d$ states of Pd, but at the Fermi level the DOS is equally contributed by these states and the $p$ states of As (see Table \ref{tab:pdos} for the partial contributions of atoms to the total DOS), highlighting the importance of Pd-As hybridization.

\begin{table}[b]
\caption{Partial densities of states in eV$^{-1}$ contributed per each atom and summed over all equivalent atoms.}
    \label{tab:pdos}
\begin{ruledtabular}
      \begin{tabular}{cccccc}
&	total & per atom&	$s$ &	$p$	&$d$\\
\hline
Mg(8c) & 0.496	&0.124	&0.009	&0.115	&\\
Pd(2a) & 0.349	&0.349	&0.006	&0.028	&0.315\\
Pd(12d) & 2.028	&0.338	&0.015	&0.047	&0.276\\
As(12e) & 2.814	&0.469&	0.042	&0.427&	\\
    \end{tabular}
\end{ruledtabular}
\end{table}

 \begin{figure*}[t]
\includegraphics[width=\textwidth]{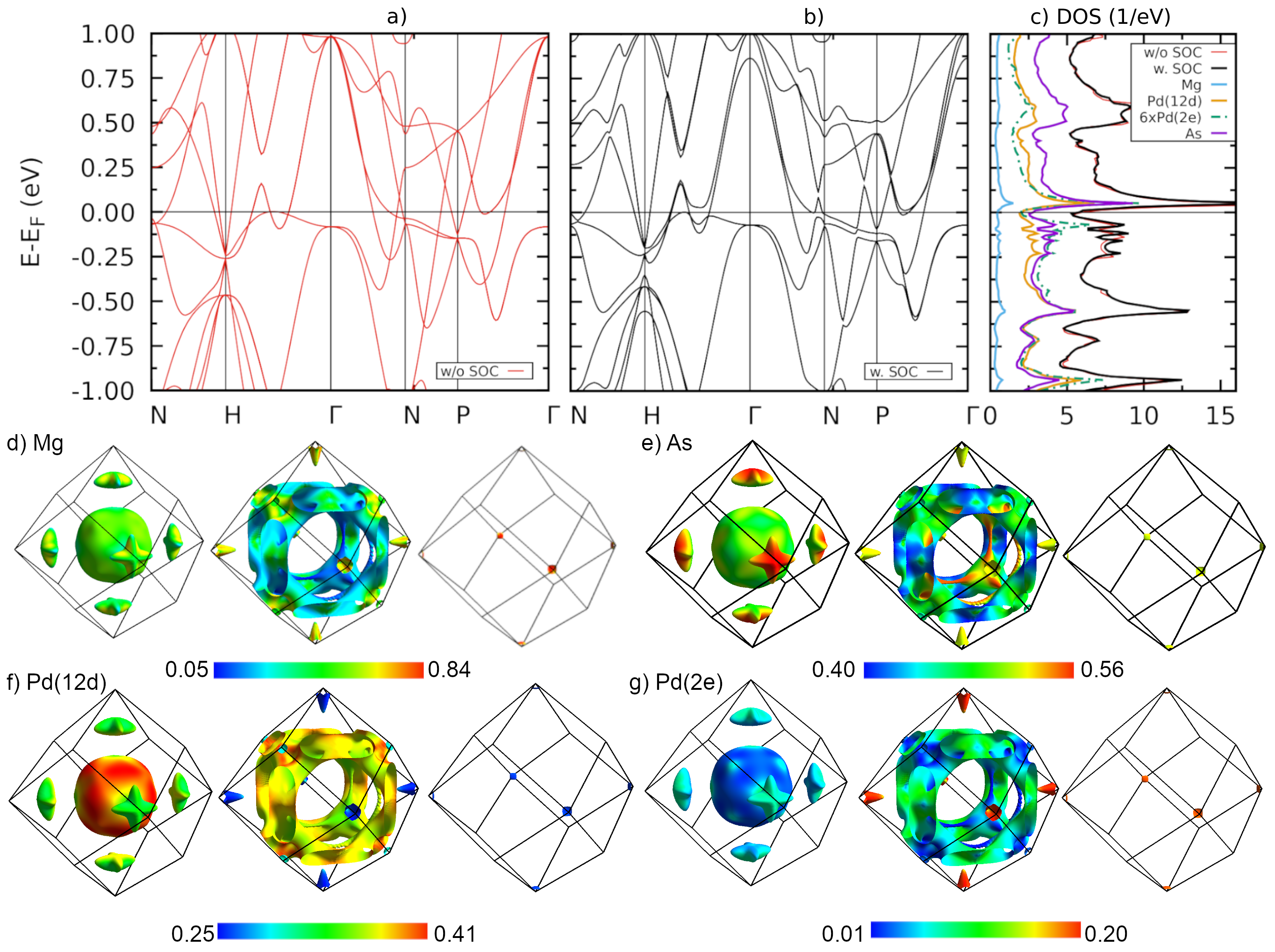} 
\caption{Electronic band structure plotted along high symmetry paths, calculated a) without SOC and b) with SOC, together with c) DOS. Furthermore, the Fermi surface is presented, colored according to the contribution of d) Mg atoms, e) As atoms,  f) Pd at the \textit{12d} site, g) Pd at the \textit{2a} site.\label{fig:band}}
 \end{figure*}

Recently, in a series of tetragonal $A$Pd$_2$P$_2$ ($A$=Cr, Sr) compounds \cite{apd2p2} superconductivity with $T_c\sim 1$ K was observed and it was suggested that short-distant Pd-P anti-bonding interactions are responsible for superconductivity. 
In this tetragonal structure, each of the Pd atoms is bonded to four P atoms, which form a tetrahedron. 
Equivalently, each P atom is bonded to four Pd atoms, forming a bypiramid. 
Similarly to the case of $A$Pd$_2$P$_2$, the crystal structure of Mg$_4$Pd$_7$As$_6$ is characterized by deformed AsPd$_4$ bypiramids, however, it does not lead to As$_4$ tetrahedrons surrounding the Pd atoms. Additionally,  there is another Pd site around which As$_6$ octahedra are formed. This provides an opportunity to obtain a deeper understanding of Pd-X (X=As, P) (anti)bonding interactions.

To discuss the nature of bonds, the crystal orbital Hamilton population (COHP) was calculated for Mg$_4$Pd$_7$As$_6$ and results are presented in Fig.~\ref{fig:dos}(b). 
At the Fermi level, only the As-As pairs form the bonding states, while the rest of pairs of atoms form the antibonding states. This underlines the importance of the As$_6$ octahedra, presented in Fig.~\ref{fig:str}(b), for the stability of this compound. This result is different than in the tetragonal CaPd$_2$P$_2$, where the Sr-Pd bonds are crucial, but what is in common is that the Pd-P antibonding interactions are crucial for superconductivity, as they dominate the DOS at Fermi level.

Figure~\ref{fig:band} presents the electronic dispersion relations of the compound studied. 
As there are 96 valence electrons per formula unit, the number of occupied bands is on the order of 48, which is why we only show the zoom of the electronic band structure around the Fermi level. In panel (a), scalar-relativistic results are plotted, whereas the spin-orbit coupling (SOC) is included in calculations presented in panel (b). SOC leads to the formation of band anticrossings at many points in the Brollouin zone; however, it does not significantly affect the total density of states, as shown in panel (c).  

\begin{figure*}[t]
\includegraphics[width=\textwidth]{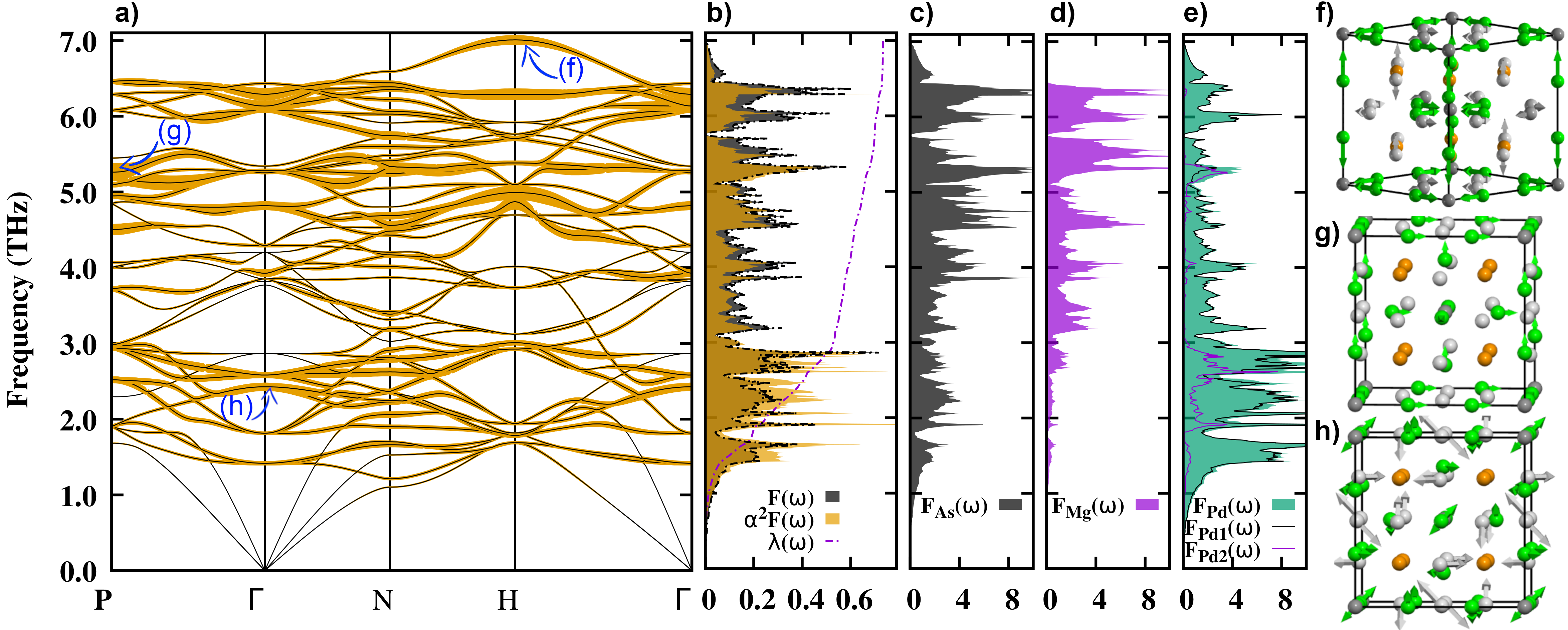} 
\caption{The phonon structure of Mg$_4$Pd$_7$As$_6$: a) phonon dispersion relations, shaded with the phonon linewidths; b) total phonon density of states $F(\omega)$, Eliashberg function $\alpha^2F(\omega)$ and cumulative electron-phonon coupling parameter $\lambda(\omega)=2\int_0^{\omega} \frac{\alpha^2F(\omega')}{\omega'} \text{d}\omega'$; partial phonon DOSes are plotted in panels c) for As; d) for Mg;  e) for Pd atoms. 
In panels (f-h) the polarization vectors associated with the phonon modes at $H$, $P$ and $\Gamma$ high symmetry points are presented. These modes are marked with arrows in panel (a).\label{fig:ph}}
\end{figure*} 
 
The Fermi level lies at the steep slope of the DOS and is crossed by three bands, giving three pieces of the Fermi surface, visualized in Fig.~\ref{fig:band}(d-g). 
Each of these panels presents the Fermi surface colored with the contribution of different atoms (Mg, As, Pd at the \textit{12d} site, Pd at the \textit{2a} site). The first piece of the Fermi surface consists of a nearly spherical $\Gamma$-centered pocket contributed equally by Pd(12d) and As states and a star-shaped pocket around, contributed mainly by As atoms, in a smaller hybridization with Pd(12d). 
The second piece consists of a large system of tubes that comes from As or Pd(12d) and small $H$-centered parabolic pockets, contributed by As and Pd(2a) atoms. The last piece consists of smaller $H$-centered pockets and comes from hybridization of the Mg, Pd(2a) and As states.  
The mean velocity of described pieces of the Fermi surface is 0.10, 0.47, 1.58$\cdot$ 10$^6$ m/s respectively, with total average 0.59 $\mu$m/s.
Contributions to $N(E_F)$ are 0.66, 2.30 and 0.004 eV$^{-1}$/spin, respectively, with a total $N(E_F)$ equal to 5.94 eV$^{-1}$.

The presence of two large Fermi surface sheets indicates the multiband nature of superconductivity in this compound. Moreover, the different orbital nature of the electronic states forming these surfaces may lead to the two-gap nature of superconductivity in Mg$_4$Pd$_7$As$_6$, which is an interesting topic for further studies.

\section{Phonons, electron-phonon interaction and superconductivity}

The phonon dispersion relations of Mg$_4$Pd$_7$As$_6$, with shading corresponding to the phonon linewidths, are shown in Fig.~\ref{fig:ph}(a). There are 17 atoms in the primitive cell; thus, there are 51 phonon modes, and the phonon spectrum is rather complicated. 
The phonon density of states, $F(\omega)$ and the Eliashberg function $\alpha^2F(\omega)$ are presented in Figs.~\ref{fig:ph}(b-e).
Comparison of the calculated lattice specific heat with the experiment, discussed above in Fig.~\ref{fig:exp-cv1}, shows good agreement, giving credit to the calculated phonon spectrum.
From the partial densities of states presented in panels (c-e), we can observe that there is a low-frequency part (up to 3 THz) dominated by Pd with smaller contribution from As atoms, whereas the rest of the spectrum is contributed by all the atoms. 

The average frequency, defined conventionally as
\begin{equation}\label{eq:sred}
\langle \omega \rangle = \int_0^{\omega_{\mathsf{max}}} \omega F(\omega) d\omega \left/ \int_0^{\omega_{\mathsf{max}}} F(\omega) d\omega \right.,
\end{equation}
is equal to 3.98 THz. 
We have checked that it is not changed by spin-orbit coupling despite changes in the band structure (we show here the results obtained with the inclusion of the SOC effect, although the scalar-relativistic results are almost the same as shown in the Supplementary Material). The average frequencies of the independent atoms are equal to 4.79 THz, 4.37 THz, 3.19 THz, and 3.08 THz for Mg, As, Pd(12d) and Pd(12a), respectively.  

The Mg atoms have the highest average frequency as a result of their smallest mass. However, it should be noted that the highest mode, appearing around the $H$ point with a frequency of approximately 7 THz, comes from the vibrations of the Pd (12d) and As atoms. 
This is the so-called breathing mode, frequently observed, e.g. in the Laves phases, where the mode is associated with a movement of atoms toward the center of close-packed tetrahedron \cite{paolasini1998lattice,gutowska2021strong,gornicka2023superconductivity}. 
The polarization vectors of this mode in Mg$_4$Pd$_7$As$_6$ are shown in Fig.~\ref{fig:ph}(f). It is associated with a displacement of As atoms toward the Pd(2a) atom and toward the center of the base of the Pd$_3$As pyramid. At the same time, the Pd(12d) atoms move towards this base. This leads to a stretching of the pyramid. Such a movement costs a large amount of energy, leading to a high frequency.

The phonon linewidth of the mode $\nu$ with wavevector $\bm q$ and frequency $\omega_{{\bf q}\nu}$ is calculated by the integration over the Fermi surface of electron-phonon interaction matrix elements $g_{{\bf q}\nu}({\bf k},i,j)$ between the electronic states $i$ with wavevector $\bm k$ and $j$ with wavevector $\bm k+\bm q$ of energy near the Fermi level
  \cite{wierzbowska,grimvall,gustino-rmp}:
\begin{equation}
\begin{split}
\gamma_{{\bf q}\nu} =& 2\pi\omega_{{\bf q}\nu} \sum_{ij}
                \int {\frac{d^3k}{ \Omega_{\rm BZ}}}  |g_{{\bf q}\nu}({\bf k},i,j)|^2 \\
                    &\times\delta(E_{{\bf k},i} - E_F)  \delta(E_{{\bf k+q},j} - E_F).
\end{split}
\end{equation}

The largest phonon linewidths are observed for the phonon mode at $P$ $\bm q$-point, marked in Fig.~\ref{fig:ph}(a) with a blue arrow. The polarization vectors of this mode are visualized in panel (g) and show that it is associated with the movement of As atoms along the $x,~y,~z$ directions. This mode disturbs the strongest bonds in the compound (Pd-As bonds), leading to a strong electron-phonon coupling. 

The Eliashberg function is defined as
\cite{wierzbowska,grimvall,gustino-rmp}:
\begin{equation}
\label{eq:a2f}
\alpha^2F(\omega) = {1\over 2\pi N(E_F)}\sum_{{\bf q}\nu} 
                    \delta(\omega-\omega_{{\bf q}\nu})
                    {\gamma_{{\bf q}\nu}\over\hbar\omega_{{\bf q}\nu}}.
\end{equation}
It is strongly enhanced in the low-frequency part, because of the strong coupling of electrons to the Pd modes and division by frequency in the formula above. 
An example of a mode in this part of the spectrum that has a relatively large phonon linewidth is marked with a blue arrow at the $\Gamma$ $\bm q$-point in panel (a). Its polarization vectors are visualized in panel (h). 
This mode is associated with the movements of both the Pd and As atoms toward the Mg atoms. In this way, not only the Pd-As bonds are stretched, leading to strong electron-phonon coupling but also such a movement toward a distant Mg atom lowers the frequency of vibration, which further enhances the Eliashberg function. 

The total electron-phonon coupling constant $\lambda$ is calculated as
 \cite{grimvall}
\begin{equation}\label{eq:lam2}
\lambda=2\int_0^{\omega_{\rm max}} \frac{\alpha^2F(\omega)}{\omega} \text{d}\omega,
\end{equation}  
 and is equal to 0.76. 
This is in very good agreement with the estimate done using the McMillan formula (0.72) and shows that Mg$_4$Pd$_7$As$_6$ is moderately-coupled electron-phonon superconductor. 
The low-frequency modes, dominated by the Pd and As vibrations with frequencies up to 3 THz, contribute 70\% of the total $\lambda$, as shown in the $\lambda(\omega)$ plot in Fig. \ref{fig:ph}(b).

\begin{figure}[t] 
\includegraphics[width=.5\textwidth]{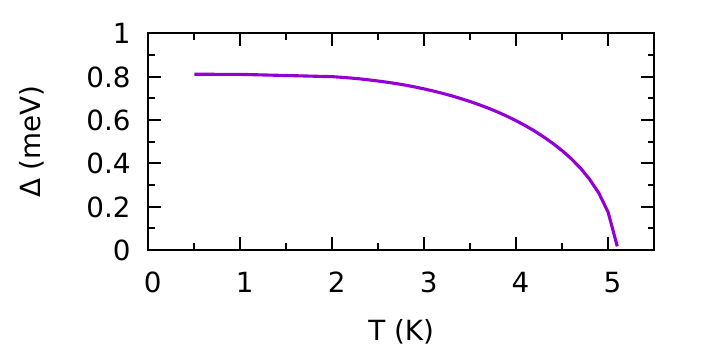} 
\caption{Superconducting gap of Mg$_4$Pd$_7$As$_6$ determined by solving the Eliashberg equations on the basis of the calculated Eliashberg function.\label{fig:gap}}
 \end{figure}

The critical temperature may be now calculated with help of Allen-Dynes equation \cite{allen-dynes}:
\begin{equation}
T_c=\frac{\omega_{\rm ln}}{1.2}
\exp\left( 
\frac{-1.04(1+\lambda)}{\lambda-\mu^*(1+0.62\lambda)}\right), 
\label{eq:tc}
\end{equation}
where the characteristic phonon frequency is defined as
\begin{equation}\label{eq:omlog2}
\omega_{\rm ln} = \exp\left(\frac{2}{\lambda}\int_0^{\omega_{\mathsf{max}}} \alpha^2F(\omega) \ln\omega\frac{d\omega}{{\omega}} \right).
\end{equation}
The resulting $T_c$ is equal to 4.95 K if a Coulomb pseudopotential value of $\mu^*=0.10$ is used and is in good agreement with the experimental result of $T_c=5.45$ K.

The agreement is even better if instead of the Allen-Dynes formula, the isotropic Eliashberg gap equations \cite{eliashberg1960interactions,PhysRevB.87.024505} are solved. In this way, the actual frequency dependence of the electron-phonon interaction strength is taken into account in the superconducting gap $\Delta(T)$ calculations. $T_c$ is determined as the temperature at which the gap vanishes. 
Eliashberg equations were solved using the Matsubara cutoff frequency $\omega_c$, which is 14 times higher than the maximal phonon frequency $\omega_{\rm max}$. The Coulomb pseudopotential parameter for Eliashberg equations was rescaled from $\mu^*=0.10$, according to the Allen and Dynes \cite{allen-dynes} method, to $\mu^*_C=\frac{1}{
\frac{1}{\mu^*}-\ln(\frac{\omega_c}{\omega_{\rm max}})}=0.14$.
The temperature dependence of the superconducting gap, obtained from the Eliashberg equations, is shown in Fig.~\ref{fig:gap}. The gap is zero at 5.1 K, closer to the experimental value. 
The gap at 0 K is equal to 0.81 meV, giving a characteristic ratio of $2\Delta/k_BT_c = 3.69$, slightly higher than the BCS prediction of 3.52.

\begin{table}[t]
\caption{The electronic density of states, average phonon frequency, electron-phonon coupling constant, logarithmic average frequency  and critical temperature of Mg$_4$Pd$_7$As$_6$ calculated using the Allen-Dynes formula ($T_c^{\rm AD}$) and Eliashberg equations ($T_c^{\rm E}$).\label{tab:tc}}
 \begin{ruledtabular}
 \begin{tabular}{ c c c c c c }
 $N(E_F)$ &	$\langle\omega\rangle$ 	&	$\lambda$	&	$\omega_{\rm ln}$ 	&	$T_c^{\rm AD}$ & $T_c^{\rm E}$	\\
 (eV$^{-1}$)& (THz) & & (K) &(K) & (K)\\
\hline
5.94	&	3.98&	0.76	&	117.83	&	4.95 & 5.1	
 \end{tabular}
 \end{ruledtabular}
 \end{table}

The agreement between the calculated and experimental superconducting parameters, $\lambda$ and $T_c$, is very good. The only disagreement between theory and experiment is in the Sommerfeld parameter, as the experimental value is $\gamma_{\rm expt.} = 48(1)$ mJ/(mol K$^2$), whereas the calculated $\gamma_{\rm calc.} = \frac{\pi^2}{3}k_B^2N(E_F)(1+\lambda)$ = 24.6 mJ/(mol K$^2$) is higher, than the calculated one. As $E_F$ is located in a local minimum of DOS function, deviations from the perfect stoichiometry may lead to the increase in $N(E_F)$. The obtained value of $\frac{\Delta C_p}{\gamma T_c} = 1.27$, which is below the BCS limit, indicates possibility of an overestimation of $\gamma$. This subject requires further analysis.

\section{Summary and conclusions}

In this work, the synthesis, crystallographic, electronic transport, magnetic and thermal properties of a new ternary Mg$_4$Pd$_7$As$_6$ compound were studied. Mg$_4$Pd$_7$As$_6$ was synthesized by solid-state reaction in a sealed quartz ampule. Crystallographic studies by means of Rietveld analysis show that this compound forms in a cubic U$_4$Re$_7$Si$_6$ structure type ($Im$-$3m$, space group No. 229). It is worth noting that Mg$_4$Pd$_7$As$_6$ is the first ternary compound in the Mg-Pd-As system.

Temperature-dependent magnetic susceptibility, electrical resistivity, and specific heat measurements confirm bulk superconductivity with the critical temperature $T_c$ = 5.5 K. The isothermal field-dependent magnetization reveals type II superconductivity and the lower critical field was estimated to be $\mu_0H_{c1}(0) = 26.7$ mT. Resistivity measurements performed at different magnetic fields in the superconducting state give the upper critical field $\mu_0H_{c2}(0) = 2.09$ T. 
Knowing the values of both critical fields, the penetration depth $\lambda_{GL}$ and the coherence length $\xi_{GL}$ were estimated, and finally the Ginzburg-Landau parameter was obtained $\kappa_{GL} = 9.4\gg \frac{1}{\sqrt{2}}$, confirming that Mg$_4$Pd$_7$As$_6$ is a type II superconductor.
Detailed studies of the specific heat data in the normal state give the Sommerfeld parameter $\gamma = 48(1)~\frac{\rm mJ}{{\rm mol} \cdot {\rm K}^2}$ and the Debye temperature of 228 K. The measurement performed in the zero field shows a clear specific heat jump $\frac{\Delta C_p}{ \gamma T_c}$ = 1.27. We also calculated the electron-phonon coupling constant using the inverted McMillan’s relation and obtained a value of $\lambda = 0.72$, indicating moderate coupling. 

Electronic structure calculations show that the Palladium $d$ states and Arsenic $p$ states dominate near the Fermi level and the anti-bonding nature of their interaction is crucial to determine the physical properties of this compound. The strongest electron-phonon coupling is observed for Pd phonon modes that distort the Pd-As tetrahedra. The calculated total electron-phonon coupling parameter of $\lambda = 0.76$ and the superconducting critical temperature $T_c = 5.1$~K agree well with the experimental findings that confirm the conventional superconducting pairing in this structure.

It is also worth mentioning that there is a cube-truncated octahedron in the crystal structures of Mg$_4$Pd$_7$As$_6$ and Mg$_{10}$Ir$_{19}$B$_{16}$, formed by Pd and Ir atoms, respectively. The presence of this crystallographic block may therefore be important for the observed superconductivity.

To the best of our knowledge, Mg$_4$Pd$_7$As$_6$ is the first reported superconductor in the 4-7-6 family. A review of the literature clearly shows that most of the reported 4:7:6 compounds contain rare earths or, less commonly, actinides. The three magnesium-containing compounds  are the only exceptions so far. Since no physical properties of these compounds have been investigated, the 4:7:6 intermetallics containing alkaline earth metals remain open for further investigation.

\section*{Acknowledgements}
The work in Gdansk was supported by National Science Centre (Poland), Grant no. 2022/45/B/ST5/03916. BW was supported by the National Science Centre (Poland), project no. 2017/26/E/ST3/00119.
SG and BW gratefully acknowledge the Polish High-Performance Computing Infrastructure PLGrid (HPC Center: ACK Cyfronet AGH) for providing computer facilities and support within computational grant no. PLG/2023/016451.

\bibliography{bib}

\begin{thebibliography}{36}%
\makeatletter
\providecommand \@ifxundefined [1]{%
 \@ifx{#1\undefined}
}%
\providecommand \@ifnum [1]{%
 \ifnum #1\expandafter \@firstoftwo
 \else \expandafter \@secondoftwo
 \fi
}%
\providecommand \@ifx [1]{%
 \ifx #1\expandafter \@firstoftwo
 \else \expandafter \@secondoftwo
 \fi
}%
\providecommand \natexlab [1]{#1}%
\providecommand \enquote  [1]{``#1''}%
\providecommand \bibnamefont  [1]{#1}%
\providecommand \bibfnamefont [1]{#1}%
\providecommand \citenamefont [1]{#1}%
\providecommand \href@noop [0]{\@secondoftwo}%
\providecommand \href [0]{\begingroup \@sanitize@url \@href}%
\providecommand \@href[1]{\@@startlink{#1}\@@href}%
\providecommand \@@href[1]{\endgroup#1\@@endlink}%
\providecommand \@sanitize@url [0]{\catcode `\\12\catcode `\$12\catcode `\&12\catcode `\#12\catcode `\^12\catcode `\_12\catcode `\%12\relax}%
\providecommand \@@startlink[1]{}%
\providecommand \@@endlink[0]{}%
\providecommand \url  [0]{\begingroup\@sanitize@url \@url }%
\providecommand \@url [1]{\endgroup\@href {#1}{\urlprefix }}%
\providecommand \urlprefix  [0]{URL }%
\providecommand \Eprint [0]{\href }%
\providecommand \doibase [0]{http://dx.doi.org/}%
\providecommand \selectlanguage [0]{\@gobble}%
\providecommand \bibinfo  [0]{\@secondoftwo}%
\providecommand \bibfield  [0]{\@secondoftwo}%
\providecommand \translation [1]{[#1]}%
\providecommand \BibitemOpen [0]{}%
\providecommand \bibitemStop [0]{}%
\providecommand \bibitemNoStop [0]{.\EOS\space}%
\providecommand \EOS [0]{\spacefactor3000\relax}%
\providecommand \BibitemShut  [1]{\csname bibitem#1\endcsname}%
\let\auto@bib@innerbib\@empty
\bibitem [{\citenamefont {Anand}\ \emph {et~al.}(2013)\citenamefont {Anand}, \citenamefont {Kim}, \citenamefont {Tanatar}, \citenamefont {Prozorov},\ and\ \citenamefont {Johnston}}]{apd2as2}%
  \BibitemOpen
  \bibfield  {author} {\bibinfo {author} {\bibfnamefont {VK}~\bibnamefont {Anand}}, \bibinfo {author} {\bibfnamefont {H}~\bibnamefont {Kim}}, \bibinfo {author} {\bibfnamefont {MA}~\bibnamefont {Tanatar}}, \bibinfo {author} {\bibfnamefont {R}~\bibnamefont {Prozorov}}, \ and\ \bibinfo {author} {\bibfnamefont {DC}~\bibnamefont {Johnston}},\ }\bibfield  {title} {\enquote {\bibinfo {title} {Superconducting and normal-state properties of {APd}$_2${As}$_2$ ({A= Ca, Sr, Ba}) single crystals},}\ }\href@noop {} {\bibfield  {journal} {\bibinfo  {journal} {Physical Review B}\ }\textbf {\bibinfo {volume} {87}},\ \bibinfo {pages} {224510} (\bibinfo {year} {2013})}\BibitemShut {NoStop}%
\bibitem [{\citenamefont {Raub}\ and\ \citenamefont {Webb}(1963)}]{pd-as}%
  \BibitemOpen
  \bibfield  {author} {\bibinfo {author} {\bibfnamefont {Ch~J}\ \bibnamefont {Raub}}\ and\ \bibinfo {author} {\bibfnamefont {GW}~\bibnamefont {Webb}},\ }\bibfield  {title} {\enquote {\bibinfo {title} {An investigation of the phase-diagram palladium-arsenic in connection with superconductivity},}\ }\href@noop {} {\bibfield  {journal} {\bibinfo  {journal} {Journal of the Less Common Metals}\ }\textbf {\bibinfo {volume} {5}},\ \bibinfo {pages} {271--277} (\bibinfo {year} {1963})}\BibitemShut {NoStop}%
\bibitem [{\citenamefont {Winiarski}\ \emph {et~al.}(2021)\citenamefont {Winiarski}, \citenamefont {Kuderowicz}, \citenamefont {G\'ornicka}, \citenamefont {Litzbarski}, \citenamefont {Stolecka}, \citenamefont {Wiendlocha}, \citenamefont {Cava},\ and\ \citenamefont {Klimczuk}}]{mgpd2sb}%
  \BibitemOpen
  \bibfield  {author} {\bibinfo {author} {\bibfnamefont {M.~J.}\ \bibnamefont {Winiarski}}, \bibinfo {author} {\bibfnamefont {G.}~\bibnamefont {Kuderowicz}}, \bibinfo {author} {\bibfnamefont {K.}~\bibnamefont {G\'ornicka}}, \bibinfo {author} {\bibfnamefont {L.~S.}\ \bibnamefont {Litzbarski}}, \bibinfo {author} {\bibfnamefont {K.}~\bibnamefont {Stolecka}}, \bibinfo {author} {\bibfnamefont {B.}~\bibnamefont {Wiendlocha}}, \bibinfo {author} {\bibfnamefont {R.~J.}\ \bibnamefont {Cava}}, \ and\ \bibinfo {author} {\bibfnamefont {T.}~\bibnamefont {Klimczuk}},\ }\bibfield  {title} {\enquote {\bibinfo {title} {{Mg}{Pd}$_{2}${Sb}: A {M}g-based {H}eusler-type superconductor},}\ }\href {\doibase 10.1103/PhysRevB.103.214501} {\bibfield  {journal} {\bibinfo  {journal} {Phys. Rev. B}\ }\textbf {\bibinfo {volume} {103}},\ \bibinfo {pages} {214501} (\bibinfo {year} {2021})}\BibitemShut {NoStop}%
\bibitem [{\citenamefont {Winiarski}\ \emph {et~al.}(2022)\citenamefont {Winiarski}, \citenamefont {Stolecka}, \citenamefont {Litzbarski}, \citenamefont {Tran}, \citenamefont {G{\'o}rnicka},\ and\ \citenamefont {Klimczuk}}]{mgpdsb}%
  \BibitemOpen
  \bibfield  {author} {\bibinfo {author} {\bibfnamefont {Micha{\l}~J}\ \bibnamefont {Winiarski}}, \bibinfo {author} {\bibfnamefont {Kamila}\ \bibnamefont {Stolecka}}, \bibinfo {author} {\bibfnamefont {Leszek}\ \bibnamefont {Litzbarski}}, \bibinfo {author} {\bibfnamefont {Thao~T}\ \bibnamefont {Tran}}, \bibinfo {author} {\bibfnamefont {Karolina}\ \bibnamefont {G{\'o}rnicka}}, \ and\ \bibinfo {author} {\bibfnamefont {Tomasz}\ \bibnamefont {Klimczuk}},\ }\bibfield  {title} {\enquote {\bibinfo {title} {{MgPdSb} - an electron-deficient half-{H}eusler phase},}\ }\href@noop {} {\bibfield  {journal} {\bibinfo  {journal} {The Journal of Physical Chemistry C}\ }\textbf {\bibinfo {volume} {126}},\ \bibinfo {pages} {14229--14235} (\bibinfo {year} {2022})}\BibitemShut {NoStop}%
\bibitem [{\citenamefont {Akselrud}\ \emph {et~al.}(1978)\citenamefont {Akselrud}, \citenamefont {Jarmoljuk},\ and\ \citenamefont {Gladyshevskij}}]{akselrud1978-U4Re7Si6}%
  \BibitemOpen
  \bibfield  {author} {\bibinfo {author} {\bibfnamefont {LG}~\bibnamefont {Akselrud}}, \bibinfo {author} {\bibfnamefont {Ja~P}\ \bibnamefont {Jarmoljuk}}, \ and\ \bibinfo {author} {\bibfnamefont {EI}~\bibnamefont {Gladyshevskij}},\ }\bibfield  {title} {\enquote {\bibinfo {title} {Crystal-structures of {U}$_4${R}e$_7${S}i$_6$ and {U}$_4$({Re}$_{0. 17}${Si}$_{0. 83}$)$_{13}$},}\ }\href@noop {} {\bibfield  {journal} {\bibinfo  {journal} {DOPOVIDI AKADEMII NAUK UKRAINSKOI RSR SERIYA A-FIZIKO-MATEMATICHNI TA TECHNICHNI NAUKI}\ ,\ \bibinfo {pages} {359--362}} (\bibinfo {year} {1978})}\BibitemShut {NoStop}%
\bibitem [{\citenamefont {Noël}\ \emph {et~al.}(2000)\citenamefont {Noël}, \citenamefont {Potel},\ and\ \citenamefont {Kaczorowski}}]{NOEL2000-U4Ru7As6}%
  \BibitemOpen
  \bibfield  {author} {\bibinfo {author} {\bibfnamefont {H}~\bibnamefont {Noël}}, \bibinfo {author} {\bibfnamefont {M}~\bibnamefont {Potel}}, \ and\ \bibinfo {author} {\bibfnamefont {D}~\bibnamefont {Kaczorowski}},\ }\bibfield  {title} {\enquote {\bibinfo {title} {A new ternary uranium arsenide, {U}$_4${R}u$_7${As}$_6$},}\ }\href {\doibase https://doi.org/10.1016/S0925-8388(99)00584-8} {\bibfield  {journal} {\bibinfo  {journal} {Journal of Alloys and Compounds}\ }\textbf {\bibinfo {volume} {302}},\ \bibinfo {pages} {L1--L2} (\bibinfo {year} {2000})}\BibitemShut {NoStop}%
\bibitem [{\citenamefont {Wurth}\ \emph {et~al.}(2001{\natexlab{a}})\citenamefont {Wurth}, \citenamefont {Löhken},\ and\ \citenamefont {Mewis}}]{Wurth_Mg}%
  \BibitemOpen
  \bibfield  {author} {\bibinfo {author} {\bibfnamefont {A.}~\bibnamefont {Wurth}}, \bibinfo {author} {\bibfnamefont {A.}~\bibnamefont {Löhken}}, \ and\ \bibinfo {author} {\bibfnamefont {A.}~\bibnamefont {Mewis}},\ }\bibfield  {title} {\enquote {\bibinfo {title} {Neue ternäre {R}hodium- und {I}ridium-{P}hosphide und -{A}rsenide mit {U}$_4${R}e$_7${S}i$_6$-struktur},}\ }\href {\doibase https://doi.org/10.1002/1521-3749(200106)627:6<1213::AID-ZAAC1213>3.0.CO;2-Z} {\bibfield  {journal} {\bibinfo  {journal} {Zeitschrift für anorganische und allgemeine Chemie}\ }\textbf {\bibinfo {volume} {627}},\ \bibinfo {pages} {1213--1216} (\bibinfo {year} {2001}{\natexlab{a}})}\BibitemShut {NoStop}%
\bibitem [{\citenamefont {Hirose}\ \emph {et~al.}(2022)\citenamefont {Hirose}, \citenamefont {Arakawa}, \citenamefont {Kato}, \citenamefont {Uwatoko}, \citenamefont {Ma}, \citenamefont {Gouchi}, \citenamefont {Honda},\ and\ \citenamefont {Settai}}]{HIROSE-Yb4Ru7As6}%
  \BibitemOpen
  \bibfield  {author} {\bibinfo {author} {\bibfnamefont {Yusuke}\ \bibnamefont {Hirose}}, \bibinfo {author} {\bibfnamefont {Kyoma}\ \bibnamefont {Arakawa}}, \bibinfo {author} {\bibfnamefont {Yuta}\ \bibnamefont {Kato}}, \bibinfo {author} {\bibfnamefont {Yoshiya}\ \bibnamefont {Uwatoko}}, \bibinfo {author} {\bibfnamefont {Hanming}\ \bibnamefont {Ma}}, \bibinfo {author} {\bibfnamefont {Jun}\ \bibnamefont {Gouchi}}, \bibinfo {author} {\bibfnamefont {Fuminori}\ \bibnamefont {Honda}}, \ and\ \bibinfo {author} {\bibfnamefont {Rikio}\ \bibnamefont {Settai}},\ }\bibfield  {title} {\enquote {\bibinfo {title} {Antiferromagnetic order in {Yb}$_4${Ru}$_7${As}$_6$ with the cubic {U}$_4${R}e$_7${S}i$_6$-type structure},}\ }\href {\doibase https://doi.org/10.1016/j.jmmm.2022.169327} {\bibfield  {journal} {\bibinfo  {journal} {Journal of Magnetism and Magnetic Materials}\ }\textbf {\bibinfo {volume} {556}},\ \bibinfo {pages} {169327} (\bibinfo {year} {2022})}\BibitemShut {NoStop}%
\bibitem [{\citenamefont {Schellenberg}\ \emph {et~al.}(2013)\citenamefont {Schellenberg}, \citenamefont {Rodewald}, \citenamefont {Schwickert}, \citenamefont {Eul},\ and\ \citenamefont {Pöttgen}}]{Schellenberg-antimonides}%
  \BibitemOpen
  \bibfield  {author} {\bibinfo {author} {\bibfnamefont {Inga}\ \bibnamefont {Schellenberg}}, \bibinfo {author} {\bibfnamefont {Ute~Ch.}\ \bibnamefont {Rodewald}}, \bibinfo {author} {\bibfnamefont {Christian}\ \bibnamefont {Schwickert}}, \bibinfo {author} {\bibfnamefont {Matthias}\ \bibnamefont {Eul}}, \ and\ \bibinfo {author} {\bibfnamefont {Rainer}\ \bibnamefont {Pöttgen}},\ }\bibfield  {title} {\enquote {\bibinfo {title} {Ternary antimonides {RE}$_4${T}$_7${Sb}$_6$ ({RE=Gd–Lu; T =Ru, Rh}) with cubic {U}$_4${R}e$_7${S}i$_6$-type structure},}\ }\href {\doibase doi:10.5560/znb.2013-3181} {\bibfield  {journal} {\bibinfo  {journal} {Zeitschrift für Naturforschung B}\ }\textbf {\bibinfo {volume} {68}},\ \bibinfo {pages} {971--978} (\bibinfo {year} {2013})}\BibitemShut {NoStop}%
\bibitem [{\citenamefont {Toby}\ and\ \citenamefont {Von~Dreele}(2013)}]{Toby:aj5212}%
  \BibitemOpen
  \bibfield  {author} {\bibinfo {author} {\bibfnamefont {Brian~H.}\ \bibnamefont {Toby}}\ and\ \bibinfo {author} {\bibfnamefont {Robert~B.}\ \bibnamefont {Von~Dreele}},\ }\bibfield  {title} {\enquote {\bibinfo {title} {{{\it GSAS-II}: the genesis of a modern open-source all purpose crystallography software package}},}\ }\href {\doibase 10.1107/S0021889813003531} {\bibfield  {journal} {\bibinfo  {journal} {Journal of Applied Crystallography}\ }\textbf {\bibinfo {volume} {46}},\ \bibinfo {pages} {544--549} (\bibinfo {year} {2013})}\BibitemShut {NoStop}%
\bibitem [{\citenamefont {Giannozzi}\ \emph {et~al.}(2009)\citenamefont {Giannozzi}, \citenamefont {Baroni}, \citenamefont {Bonini}, \citenamefont {Calandra}, \citenamefont {Car}, \citenamefont {Cavazzoni}, \citenamefont {Ceresoli}, \citenamefont {Chiarotti}, \citenamefont {Cococcioni}, \citenamefont {Dabo}, \citenamefont {{Dal Corso}}, \citenamefont {de~Gironcoli}, \citenamefont {Fabris}, \citenamefont {Fratesi}, \citenamefont {Gebauer}, \citenamefont {Gerstmann}, \citenamefont {Gougoussis}, \citenamefont {Kokalj}, \citenamefont {Lazzeri}, \citenamefont {Martin-Samos}, \citenamefont {Marzari}, \citenamefont {Mauri}, \citenamefont {Mazzarello}, \citenamefont {Paolini}, \citenamefont {Pasquarello}, \citenamefont {Paulatto}, \citenamefont {Sbraccia}, \citenamefont {Scandolo}, \citenamefont {Sclauzero}, \citenamefont {Seitsonen}, \citenamefont {Smogunov}, \citenamefont {Umari},\ and\ \citenamefont {Wentzcovitch}}]{qe}%
  \BibitemOpen
  \bibfield  {author} {\bibinfo {author} {\bibfnamefont {Paolo}\ \bibnamefont {Giannozzi}}, \bibinfo {author} {\bibfnamefont {Stefano}\ \bibnamefont {Baroni}}, \bibinfo {author} {\bibfnamefont {Nicola}\ \bibnamefont {Bonini}}, \bibinfo {author} {\bibfnamefont {Matteo}\ \bibnamefont {Calandra}}, \bibinfo {author} {\bibfnamefont {Roberto}\ \bibnamefont {Car}}, \bibinfo {author} {\bibfnamefont {Carlo}\ \bibnamefont {Cavazzoni}}, \bibinfo {author} {\bibfnamefont {Davide}\ \bibnamefont {Ceresoli}}, \bibinfo {author} {\bibfnamefont {Guido~L}\ \bibnamefont {Chiarotti}}, \bibinfo {author} {\bibfnamefont {Matteo}\ \bibnamefont {Cococcioni}}, \bibinfo {author} {\bibfnamefont {Ismaila}\ \bibnamefont {Dabo}}, \bibinfo {author} {\bibfnamefont {Andrea}\ \bibnamefont {{Dal Corso}}}, \bibinfo {author} {\bibfnamefont {Stefano}\ \bibnamefont {de~Gironcoli}}, \bibinfo {author} {\bibfnamefont {Stefano}\ \bibnamefont {Fabris}}, \bibinfo {author} {\bibfnamefont {Guido}\ \bibnamefont {Fratesi}}, \bibinfo {author} {\bibfnamefont
  {Ralph}\ \bibnamefont {Gebauer}}, \bibinfo {author} {\bibfnamefont {Uwe}\ \bibnamefont {Gerstmann}}, \bibinfo {author} {\bibfnamefont {Christos}\ \bibnamefont {Gougoussis}}, \bibinfo {author} {\bibfnamefont {Anton}\ \bibnamefont {Kokalj}}, \bibinfo {author} {\bibfnamefont {Michele}\ \bibnamefont {Lazzeri}}, \bibinfo {author} {\bibfnamefont {Layla}\ \bibnamefont {Martin-Samos}}, \bibinfo {author} {\bibfnamefont {Nicola}\ \bibnamefont {Marzari}}, \bibinfo {author} {\bibfnamefont {Francesco}\ \bibnamefont {Mauri}}, \bibinfo {author} {\bibfnamefont {Riccardo}\ \bibnamefont {Mazzarello}}, \bibinfo {author} {\bibfnamefont {Stefano}\ \bibnamefont {Paolini}}, \bibinfo {author} {\bibfnamefont {Alfredo}\ \bibnamefont {Pasquarello}}, \bibinfo {author} {\bibfnamefont {Lorenzo}\ \bibnamefont {Paulatto}}, \bibinfo {author} {\bibfnamefont {Carlo}\ \bibnamefont {Sbraccia}}, \bibinfo {author} {\bibfnamefont {Sandro}\ \bibnamefont {Scandolo}}, \bibinfo {author} {\bibfnamefont {Gabriele}\ \bibnamefont {Sclauzero}}, \bibinfo
  {author} {\bibfnamefont {Ari~P}\ \bibnamefont {Seitsonen}}, \bibinfo {author} {\bibfnamefont {Alexander}\ \bibnamefont {Smogunov}}, \bibinfo {author} {\bibfnamefont {Paolo}\ \bibnamefont {Umari}}, \ and\ \bibinfo {author} {\bibfnamefont {Renata~M}\ \bibnamefont {Wentzcovitch}},\ }\bibfield  {title} {\enquote {\bibinfo {title} {{QUANTUM ESPRESSO}: a modular and open-source software project for quantum simulations of materials},}\ }\href {http://www.quantum-espresso.org} {\bibfield  {journal} {\bibinfo  {journal} {Journal of Physics: Condensed Matter}\ }\textbf {\bibinfo {volume} {21}},\ \bibinfo {pages} {395502 (19pp)} (\bibinfo {year} {2009})}\BibitemShut {NoStop}%
\bibitem [{\citenamefont {Giannozzi}\ \emph {et~al.}(2017)\citenamefont {Giannozzi}, \citenamefont {Andreussi}, \citenamefont {Brumme}, \citenamefont {Bunau}, \citenamefont {Nardelli}, \citenamefont {Calandra}, \citenamefont {Car}, \citenamefont {Cavazzoni}, \citenamefont {Ceresoli}, \citenamefont {Cococcioni}, \citenamefont {Colonna}, \citenamefont {Carnimeo}, \citenamefont {Corso}, \citenamefont {de~Gironcoli}, \citenamefont {Delugas}, \citenamefont {DiStasio}, \citenamefont {Ferretti}, \citenamefont {Floris}, \citenamefont {Fratesi}, \citenamefont {Fugallo}, \citenamefont {Gebauer}, \citenamefont {Gerstmann}, \citenamefont {Giustino}, \citenamefont {Gorni}, \citenamefont {Jia}, \citenamefont {Kawamura}, \citenamefont {Ko}, \citenamefont {Kokalj}, \citenamefont {Kü{\c{c}}ükbenli}, \citenamefont {Lazzeri}, \citenamefont {Marsili}, \citenamefont {Marzari}, \citenamefont {Mauri}, \citenamefont {Nguyen}, \citenamefont {Nguyen}, \citenamefont {de-la Roza}, \citenamefont {Paulatto}, \citenamefont {Ponc{\'{e}}},
  \citenamefont {Rocca}, \citenamefont {Sabatini}, \citenamefont {Santra}, \citenamefont {Schlipf}, \citenamefont {Seitsonen}, \citenamefont {Smogunov}, \citenamefont {Timrov}, \citenamefont {Thonhauser}, \citenamefont {Umari}, \citenamefont {Vast}, \citenamefont {Wu},\ and\ \citenamefont {Baroni}}]{qe2}%
  \BibitemOpen
  \bibfield  {author} {\bibinfo {author} {\bibfnamefont {P}~\bibnamefont {Giannozzi}}, \bibinfo {author} {\bibfnamefont {O}~\bibnamefont {Andreussi}}, \bibinfo {author} {\bibfnamefont {T}~\bibnamefont {Brumme}}, \bibinfo {author} {\bibfnamefont {O}~\bibnamefont {Bunau}}, \bibinfo {author} {\bibfnamefont {M~Buongiorno}\ \bibnamefont {Nardelli}}, \bibinfo {author} {\bibfnamefont {M}~\bibnamefont {Calandra}}, \bibinfo {author} {\bibfnamefont {R}~\bibnamefont {Car}}, \bibinfo {author} {\bibfnamefont {C}~\bibnamefont {Cavazzoni}}, \bibinfo {author} {\bibfnamefont {D}~\bibnamefont {Ceresoli}}, \bibinfo {author} {\bibfnamefont {M}~\bibnamefont {Cococcioni}}, \bibinfo {author} {\bibfnamefont {N}~\bibnamefont {Colonna}}, \bibinfo {author} {\bibfnamefont {I}~\bibnamefont {Carnimeo}}, \bibinfo {author} {\bibfnamefont {A~Dal}\ \bibnamefont {Corso}}, \bibinfo {author} {\bibfnamefont {S}~\bibnamefont {de~Gironcoli}}, \bibinfo {author} {\bibfnamefont {P}~\bibnamefont {Delugas}}, \bibinfo {author} {\bibfnamefont {R~A}\
  \bibnamefont {DiStasio}}, \bibinfo {author} {\bibfnamefont {A}~\bibnamefont {Ferretti}}, \bibinfo {author} {\bibfnamefont {A}~\bibnamefont {Floris}}, \bibinfo {author} {\bibfnamefont {G}~\bibnamefont {Fratesi}}, \bibinfo {author} {\bibfnamefont {G}~\bibnamefont {Fugallo}}, \bibinfo {author} {\bibfnamefont {R}~\bibnamefont {Gebauer}}, \bibinfo {author} {\bibfnamefont {U}~\bibnamefont {Gerstmann}}, \bibinfo {author} {\bibfnamefont {F}~\bibnamefont {Giustino}}, \bibinfo {author} {\bibfnamefont {T}~\bibnamefont {Gorni}}, \bibinfo {author} {\bibfnamefont {J}~\bibnamefont {Jia}}, \bibinfo {author} {\bibfnamefont {M}~\bibnamefont {Kawamura}}, \bibinfo {author} {\bibfnamefont {H-Y}\ \bibnamefont {Ko}}, \bibinfo {author} {\bibfnamefont {A}~\bibnamefont {Kokalj}}, \bibinfo {author} {\bibfnamefont {E}~\bibnamefont {Kü{\c{c}}ükbenli}}, \bibinfo {author} {\bibfnamefont {M}~\bibnamefont {Lazzeri}}, \bibinfo {author} {\bibfnamefont {M}~\bibnamefont {Marsili}}, \bibinfo {author} {\bibfnamefont {N}~\bibnamefont
  {Marzari}}, \bibinfo {author} {\bibfnamefont {F}~\bibnamefont {Mauri}}, \bibinfo {author} {\bibfnamefont {N~L}\ \bibnamefont {Nguyen}}, \bibinfo {author} {\bibfnamefont {H-V}\ \bibnamefont {Nguyen}}, \bibinfo {author} {\bibfnamefont {A~Otero}\ \bibnamefont {de-la Roza}}, \bibinfo {author} {\bibfnamefont {L}~\bibnamefont {Paulatto}}, \bibinfo {author} {\bibfnamefont {S}~\bibnamefont {Ponc{\'{e}}}}, \bibinfo {author} {\bibfnamefont {D}~\bibnamefont {Rocca}}, \bibinfo {author} {\bibfnamefont {R}~\bibnamefont {Sabatini}}, \bibinfo {author} {\bibfnamefont {B}~\bibnamefont {Santra}}, \bibinfo {author} {\bibfnamefont {M}~\bibnamefont {Schlipf}}, \bibinfo {author} {\bibfnamefont {A~P}\ \bibnamefont {Seitsonen}}, \bibinfo {author} {\bibfnamefont {A}~\bibnamefont {Smogunov}}, \bibinfo {author} {\bibfnamefont {I}~\bibnamefont {Timrov}}, \bibinfo {author} {\bibfnamefont {T}~\bibnamefont {Thonhauser}}, \bibinfo {author} {\bibfnamefont {P}~\bibnamefont {Umari}}, \bibinfo {author} {\bibfnamefont {N}~\bibnamefont {Vast}},
  \bibinfo {author} {\bibfnamefont {X}~\bibnamefont {Wu}}, \ and\ \bibinfo {author} {\bibfnamefont {S}~\bibnamefont {Baroni}},\ }\bibfield  {title} {\enquote {\bibinfo {title} {Advanced capabilities for materials modelling with quantum {ESPRESSO}},}\ }\href {\doibase 10.1088/1361-648x/aa8f79} {\bibfield  {journal} {\bibinfo  {journal} {Journal of Physics: Condensed Matter}\ }\textbf {\bibinfo {volume} {29}},\ \bibinfo {pages} {465901} (\bibinfo {year} {2017})}\BibitemShut {NoStop}%
\bibitem [{\citenamefont {{Dal Corso}}(2014)}]{pps}%
  \BibitemOpen
  \bibfield  {author} {\bibinfo {author} {\bibfnamefont {Andrea}\ \bibnamefont {{Dal Corso}}},\ }\bibfield  {title} {\enquote {\bibinfo {title} {{Pseudopotentials periodic table: From {H} to {P}u}},}\ }\href {\doibase https://doi.org/10.1016/j.commatsci.2014.07.043} {\bibfield  {journal} {\bibinfo  {journal} {Computational Materials Science}\ }\textbf {\bibinfo {volume} {95}},\ \bibinfo {pages} {337 -- 350} (\bibinfo {year} {2014})}\BibitemShut {NoStop}%
\bibitem [{\citenamefont {Perdew}\ \emph {et~al.}(1996)\citenamefont {Perdew}, \citenamefont {Burke},\ and\ \citenamefont {Ernzerhof}}]{pbe}%
  \BibitemOpen
  \bibfield  {author} {\bibinfo {author} {\bibfnamefont {John~P.}\ \bibnamefont {Perdew}}, \bibinfo {author} {\bibfnamefont {Kieron}\ \bibnamefont {Burke}}, \ and\ \bibinfo {author} {\bibfnamefont {Matthias}\ \bibnamefont {Ernzerhof}},\ }\bibfield  {title} {\enquote {\bibinfo {title} {{Generalized Gradient Approximation Made Simple}},}\ }\href {\doibase 10.1103/PhysRevLett.77.3865} {\bibfield  {journal} {\bibinfo  {journal} {Phys. Rev. Lett.}\ }\textbf {\bibinfo {volume} {77}},\ \bibinfo {pages} {3865--3868} (\bibinfo {year} {1996})}\BibitemShut {NoStop}%
\bibitem [{\citenamefont {Baroni}\ \emph {et~al.}(2001)\citenamefont {Baroni}, \citenamefont {de~Gironcoli}, \citenamefont {Dal~Corso},\ and\ \citenamefont {Giannozzi}}]{baroni}%
  \BibitemOpen
  \bibfield  {author} {\bibinfo {author} {\bibfnamefont {Stefano}\ \bibnamefont {Baroni}}, \bibinfo {author} {\bibfnamefont {Stefano}\ \bibnamefont {de~Gironcoli}}, \bibinfo {author} {\bibfnamefont {Andrea}\ \bibnamefont {Dal~Corso}}, \ and\ \bibinfo {author} {\bibfnamefont {Paolo}\ \bibnamefont {Giannozzi}},\ }\bibfield  {title} {\enquote {\bibinfo {title} {Phonons and related crystal properties from density-functional perturbation theory},}\ }\href {\doibase 10.1103/RevModPhys.73.515} {\bibfield  {journal} {\bibinfo  {journal} {Rev. Mod. Phys.}\ }\textbf {\bibinfo {volume} {73}},\ \bibinfo {pages} {515--562} (\bibinfo {year} {2001})}\BibitemShut {NoStop}%
\bibitem [{\citenamefont {Engel}\ \emph {et~al.}(1984)\citenamefont {Engel}, \citenamefont {Chabot},\ and\ \citenamefont {Parth{\'e}}}]{engel}%
  \BibitemOpen
  \bibfield  {author} {\bibinfo {author} {\bibfnamefont {N}~\bibnamefont {Engel}}, \bibinfo {author} {\bibfnamefont {B}~\bibnamefont {Chabot}}, \ and\ \bibinfo {author} {\bibfnamefont {E}~\bibnamefont {Parth{\'e}}},\ }\bibfield  {title} {\enquote {\bibinfo {title} {Sc$_4${T}$_7${G}e$_6$ ({T}$\equiv$ {Rh, Ir, Os, Ru}) and {S}c$_4${I}r$_7${S}i$_6$ with the cubic {U}$_4${R}e$_7${S}i$_6$-type structure},}\ }\href@noop {} {\bibfield  {journal} {\bibinfo  {journal} {Journal of the Less Common Metals}\ }\textbf {\bibinfo {volume} {96}},\ \bibinfo {pages} {291--296} (\bibinfo {year} {1984})}\BibitemShut {NoStop}%
\bibitem [{\citenamefont {Matar}\ \emph {et~al.}(2014)\citenamefont {Matar}, \citenamefont {Chevalier},\ and\ \citenamefont {P{\"o}ttgen}}]{matar}%
  \BibitemOpen
  \bibfield  {author} {\bibinfo {author} {\bibfnamefont {Samir~F}\ \bibnamefont {Matar}}, \bibinfo {author} {\bibfnamefont {Bernard}\ \bibnamefont {Chevalier}}, \ and\ \bibinfo {author} {\bibfnamefont {Rainer}\ \bibnamefont {P{\"o}ttgen}},\ }\bibfield  {title} {\enquote {\bibinfo {title} {The {U}$_4${R}e$_7${S}i$_6$ type--trends in electronic structure and chemical bonding},}\ }\href@noop {} {\bibfield  {journal} {\bibinfo  {journal} {Solid state sciences}\ }\textbf {\bibinfo {volume} {27}},\ \bibinfo {pages} {5--10} (\bibinfo {year} {2014})}\BibitemShut {NoStop}%
\bibitem [{\citenamefont {Leithe-Jasper}\ \emph {et~al.}(2014)\citenamefont {Leithe-Jasper}, \citenamefont {Cardoso-Gil}, \citenamefont {Ramlau},\ and\ \citenamefont {Burkhardt}}]{Leithe-structure}%
  \BibitemOpen
  \bibfield  {author} {\bibinfo {author} {\bibfnamefont {A.}~\bibnamefont {Leithe-Jasper}}, \bibinfo {author} {\bibfnamefont {R.}~\bibnamefont {Cardoso-Gil}}, \bibinfo {author} {\bibfnamefont {Hartmut}\ \bibnamefont {Ramlau}}, \ and\ \bibinfo {author} {\bibfnamefont {U.}~\bibnamefont {Burkhardt}},\ }\bibfield  {title} {\enquote {\bibinfo {title} {Crystal structure of tetraytterbium septarhodium hexaantimony, {Yb}$_4${Rh}$_7${Sb}$_6$},}\ }\href {\doibase 10.1524/ncrs.2006.0062} {\bibfield  {journal} {\bibinfo  {journal} {Zeitschrift für Kristallographie - New Crystal Structures}\ }\textbf {\bibinfo {volume} {221}} (\bibinfo {year} {2014}),\ 10.1524/ncrs.2006.0062}\BibitemShut {NoStop}%
\bibitem [{\citenamefont {Xue}\ \emph {et~al.}(2017)\citenamefont {Xue}, \citenamefont {Sun},\ and\ \citenamefont {Chen}}]{hybridization}%
  \BibitemOpen
  \bibfield  {author} {\bibinfo {author} {\bibfnamefont {Dongfeng}\ \bibnamefont {Xue}}, \bibinfo {author} {\bibfnamefont {Congting}\ \bibnamefont {Sun}}, \ and\ \bibinfo {author} {\bibfnamefont {Xiaoyan}\ \bibnamefont {Chen}},\ }\bibfield  {title} {\enquote {\bibinfo {title} {Hybridization: a chemical bonding nature of atoms},}\ }\href@noop {} {\bibfield  {journal} {\bibinfo  {journal} {Chinese Journal of Chemistry}\ }\textbf {\bibinfo {volume} {35}},\ \bibinfo {pages} {1452--1458} (\bibinfo {year} {2017})}\BibitemShut {NoStop}%
\bibitem [{\citenamefont {Heying}\ \emph {et~al.}(2004)\citenamefont {Heying}, \citenamefont {Katoh}, \citenamefont {Niide}, \citenamefont {Ochiai},\ and\ \citenamefont {Poettgen}}]{heying}%
  \BibitemOpen
  \bibfield  {author} {\bibinfo {author} {\bibfnamefont {Birgit}\ \bibnamefont {Heying}}, \bibinfo {author} {\bibfnamefont {Kenichi}\ \bibnamefont {Katoh}}, \bibinfo {author} {\bibfnamefont {Yuzuru}\ \bibnamefont {Niide}}, \bibinfo {author} {\bibfnamefont {Akira}\ \bibnamefont {Ochiai}}, \ and\ \bibinfo {author} {\bibfnamefont {Rainer}\ \bibnamefont {Poettgen}},\ }\bibfield  {title} {\enquote {\bibinfo {title} {Synthesis and structures of {Y}b$_4${R}h$_7${G}e$_6$ and {Y}b$_4${I}r$_7${G}e$_6$},}\ }\href@noop {} {\bibfield  {journal} {\bibinfo  {journal} {Zeitschrift f{\"u}r anorganische und allgemeine Chemie}\ }\textbf {\bibinfo {volume} {630}},\ \bibinfo {pages} {1423--1426} (\bibinfo {year} {2004})}\BibitemShut {NoStop}%
\bibitem [{\citenamefont {Xu}\ \emph {et~al.}(2009)\citenamefont {Xu}, \citenamefont {Klimczuk}, \citenamefont {Gortenmulder}, \citenamefont {Jansen}, \citenamefont {McGuire}, \citenamefont {Cava},\ and\ \citenamefont {Zandbergen}}]{xu}%
  \BibitemOpen
  \bibfield  {author} {\bibinfo {author} {\bibfnamefont {Qiang}\ \bibnamefont {Xu}}, \bibinfo {author} {\bibfnamefont {Tomasz}\ \bibnamefont {Klimczuk}}, \bibinfo {author} {\bibfnamefont {Ton}\ \bibnamefont {Gortenmulder}}, \bibinfo {author} {\bibfnamefont {Jacob}\ \bibnamefont {Jansen}}, \bibinfo {author} {\bibfnamefont {Michael~A}\ \bibnamefont {McGuire}}, \bibinfo {author} {\bibfnamefont {Robert~J}\ \bibnamefont {Cava}}, \ and\ \bibinfo {author} {\bibfnamefont {Henny~W}\ \bibnamefont {Zandbergen}},\ }\bibfield  {title} {\enquote {\bibinfo {title} {Ab initio structure determination of {M}g$_{10}${I}r$_{19}${B}$_{16}$},}\ }\href@noop {} {\bibfield  {journal} {\bibinfo  {journal} {Chemistry of Materials}\ }\textbf {\bibinfo {volume} {21}},\ \bibinfo {pages} {2499--2507} (\bibinfo {year} {2009})}\BibitemShut {NoStop}%
\bibitem [{\citenamefont {Wurth}\ \emph {et~al.}(2001{\natexlab{b}})\citenamefont {Wurth}, \citenamefont {L{\"o}hken},\ and\ \citenamefont {Mewis}}]{mg4rh7as6-mg4ir7as6}%
  \BibitemOpen
  \bibfield  {author} {\bibinfo {author} {\bibfnamefont {A}~\bibnamefont {Wurth}}, \bibinfo {author} {\bibfnamefont {A}~\bibnamefont {L{\"o}hken}}, \ and\ \bibinfo {author} {\bibfnamefont {A}~\bibnamefont {Mewis}},\ }\bibfield  {title} {\enquote {\bibinfo {title} {Neue tern{\"a}re rhodium-und iridium-phosphide und-arsenide mit {U}$_4${R}e$_7${S}i$_6$-struktur},}\ }\href@noop {} {\bibfield  {journal} {\bibinfo  {journal} {Zeitschrift f{\"u}r anorganische und allgemeine Chemie}\ }\textbf {\bibinfo {volume} {627}},\ \bibinfo {pages} {1213--1216} (\bibinfo {year} {2001}{\natexlab{b}})}\BibitemShut {NoStop}%
\bibitem [{\citenamefont {Grimvall}(1981)}]{grimvall}%
  \BibitemOpen
  \bibfield  {author} {\bibinfo {author} {\bibfnamefont {G.}~\bibnamefont {Grimvall}},\ }\href@noop {} {\emph {\bibinfo {title} {The electron-phonon interaction in metals}}}\ (\bibinfo  {publisher} {North-Holland, Amsterdam},\ \bibinfo {year} {1981})\BibitemShut {NoStop}%
\bibitem [{\citenamefont {Katoh}\ \emph {et~al.}(2004)\citenamefont {Katoh}, \citenamefont {Abe}, \citenamefont {Negishi}, \citenamefont {Terui}, \citenamefont {Niide},\ and\ \citenamefont {Ochiai}}]{KATOH-Debye}%
  \BibitemOpen
  \bibfield  {author} {\bibinfo {author} {\bibfnamefont {K}~\bibnamefont {Katoh}}, \bibinfo {author} {\bibfnamefont {H}~\bibnamefont {Abe}}, \bibinfo {author} {\bibfnamefont {D}~\bibnamefont {Negishi}}, \bibinfo {author} {\bibfnamefont {G}~\bibnamefont {Terui}}, \bibinfo {author} {\bibfnamefont {Y}~\bibnamefont {Niide}}, \ and\ \bibinfo {author} {\bibfnamefont {A}~\bibnamefont {Ochiai}},\ }\bibfield  {title} {\enquote {\bibinfo {title} {Magnetic and transport properties of {Yb}$_4${Rh}$_7${Ge}$_6$ and {Yb}$_4${Ir}$_7${Ge}$_6$},}\ }\href {\doibase https://doi.org/10.1016/j.jmmm.2004.01.075} {\bibfield  {journal} {\bibinfo  {journal} {Journal of Magnetism and Magnetic Materials}\ }\textbf {\bibinfo {volume} {279}},\ \bibinfo {pages} {118--124} (\bibinfo {year} {2004})}\BibitemShut {NoStop}%
\bibitem [{\citenamefont {Mentink}\ \emph {et~al.}(1991)\citenamefont {Mentink}, \citenamefont {Nieuwenhuys}, \citenamefont {Menovsky},\ and\ \citenamefont {Mydosh}}]{Mentink-Debye}%
  \BibitemOpen
  \bibfield  {author} {\bibinfo {author} {\bibfnamefont {S.~A.~M.}\ \bibnamefont {Mentink}}, \bibinfo {author} {\bibfnamefont {G.~J.}\ \bibnamefont {Nieuwenhuys}}, \bibinfo {author} {\bibfnamefont {A.~A.}\ \bibnamefont {Menovsky}}, \ and\ \bibinfo {author} {\bibfnamefont {J.~A.}\ \bibnamefont {Mydosh}},\ }\bibfield  {title} {\enquote {\bibinfo {title} {{Thermal, electrical, and magnetic properties of {U}$_4${R}u$_7${G}e$_6$}},}\ }\href {\doibase 10.1063/1.348942} {\bibfield  {journal} {\bibinfo  {journal} {Journal of Applied Physics}\ }\textbf {\bibinfo {volume} {69}},\ \bibinfo {pages} {5484--5486} (\bibinfo {year} {1991})}\BibitemShut {NoStop}%
\bibitem [{\citenamefont {Singh}\ \emph {et~al.}(2010)\citenamefont {Singh}, \citenamefont {Martin}, \citenamefont {Bud'ko}, \citenamefont {Ellern}, \citenamefont {Prozorov},\ and\ \citenamefont {Johnston}}]{OsB2}%
  \BibitemOpen
  \bibfield  {author} {\bibinfo {author} {\bibfnamefont {Yogesh}\ \bibnamefont {Singh}}, \bibinfo {author} {\bibfnamefont {C.}~\bibnamefont {Martin}}, \bibinfo {author} {\bibfnamefont {S.~L.}\ \bibnamefont {Bud'ko}}, \bibinfo {author} {\bibfnamefont {A.}~\bibnamefont {Ellern}}, \bibinfo {author} {\bibfnamefont {R.}~\bibnamefont {Prozorov}}, \ and\ \bibinfo {author} {\bibfnamefont {D.~C.}\ \bibnamefont {Johnston}},\ }\bibfield  {title} {\enquote {\bibinfo {title} {Multigap superconductivity and shubnikov--de haas oscillations in single crystals of the layered boride ${\text{osb}}_{2}$},}\ }\href {\doibase 10.1103/PhysRevB.82.144532} {\bibfield  {journal} {\bibinfo  {journal} {Phys. Rev. B}\ }\textbf {\bibinfo {volume} {82}},\ \bibinfo {pages} {144532} (\bibinfo {year} {2010})}\BibitemShut {NoStop}%
\bibitem [{\citenamefont {de-la Roza}\ \emph {et~al.}(2014)\citenamefont {de-la Roza}, \citenamefont {Johnson},\ and\ \citenamefont {Luaña}}]{critic}%
  \BibitemOpen
  \bibfield  {author} {\bibinfo {author} {\bibfnamefont {A.~Otero}\ \bibnamefont {de-la Roza}}, \bibinfo {author} {\bibfnamefont {Erin~R.}\ \bibnamefont {Johnson}}, \ and\ \bibinfo {author} {\bibfnamefont {Víctor}\ \bibnamefont {Luaña}},\ }\bibfield  {title} {\enquote {\bibinfo {title} {Critic2: A program for real-space analysis of quantum chemical interactions in solids},}\ }\href {\doibase https://doi.org/10.1016/j.cpc.2013.10.026} {\bibfield  {journal} {\bibinfo  {journal} {Computer Physics Communications}\ }\textbf {\bibinfo {volume} {185}},\ \bibinfo {pages} {1007 -- 1018} (\bibinfo {year} {2014})}\BibitemShut {NoStop}%
\bibitem [{\citenamefont {Blawat}\ \emph {et~al.}(2020)\citenamefont {Blawat}, \citenamefont {Swatek}, \citenamefont {Das}, \citenamefont {Kaczorowski}, \citenamefont {Jin},\ and\ \citenamefont {Xie}}]{apd2p2}%
  \BibitemOpen
  \bibfield  {author} {\bibinfo {author} {\bibfnamefont {Joanna}\ \bibnamefont {Blawat}}, \bibinfo {author} {\bibfnamefont {Przemyslaw~Wojciech}\ \bibnamefont {Swatek}}, \bibinfo {author} {\bibfnamefont {Debarchan}\ \bibnamefont {Das}}, \bibinfo {author} {\bibfnamefont {Dariusz}\ \bibnamefont {Kaczorowski}}, \bibinfo {author} {\bibfnamefont {Rongying}\ \bibnamefont {Jin}}, \ and\ \bibinfo {author} {\bibfnamefont {Weiwei}\ \bibnamefont {Xie}},\ }\bibfield  {title} {\enquote {\bibinfo {title} {{P}d-{P} antibonding interactions in $a{\mathrm{pd}}_{2}{\mathrm{p}}_{2}$ ($a=\mathrm{Ca}$ and {S}r) superconductors},}\ }\href {\doibase 10.1103/PhysRevMaterials.4.014801} {\bibfield  {journal} {\bibinfo  {journal} {Phys. Rev. Mater.}\ }\textbf {\bibinfo {volume} {4}},\ \bibinfo {pages} {014801} (\bibinfo {year} {2020})}\BibitemShut {NoStop}%
\bibitem [{\citenamefont {Paolasini}\ \emph {et~al.}(1998)\citenamefont {Paolasini}, \citenamefont {Hennion}, \citenamefont {Panchula}, \citenamefont {Myers},\ and\ \citenamefont {Canfield}}]{paolasini1998lattice}%
  \BibitemOpen
  \bibfield  {author} {\bibinfo {author} {\bibfnamefont {L}~\bibnamefont {Paolasini}}, \bibinfo {author} {\bibfnamefont {B}~\bibnamefont {Hennion}}, \bibinfo {author} {\bibfnamefont {A}~\bibnamefont {Panchula}}, \bibinfo {author} {\bibfnamefont {K}~\bibnamefont {Myers}}, \ and\ \bibinfo {author} {\bibfnamefont {Paul}\ \bibnamefont {Canfield}},\ }\bibfield  {title} {\enquote {\bibinfo {title} {Lattice dynamics of cubic {L}aves phase ferromagnets},}\ }\href@noop {} {\bibfield  {journal} {\bibinfo  {journal} {Physical Review B}\ }\textbf {\bibinfo {volume} {58}},\ \bibinfo {pages} {12125} (\bibinfo {year} {1998})}\BibitemShut {NoStop}%
\bibitem [{\citenamefont {Gutowska}\ \emph {et~al.}(2021)\citenamefont {Gutowska}, \citenamefont {G{\'o}rnicka}, \citenamefont {W{\'o}jcik}, \citenamefont {Klimczuk},\ and\ \citenamefont {Wiendlocha}}]{gutowska2021strong}%
  \BibitemOpen
  \bibfield  {author} {\bibinfo {author} {\bibfnamefont {Sylwia}\ \bibnamefont {Gutowska}}, \bibinfo {author} {\bibfnamefont {Karolina}\ \bibnamefont {G{\'o}rnicka}}, \bibinfo {author} {\bibfnamefont {Pawe{\l}}\ \bibnamefont {W{\'o}jcik}}, \bibinfo {author} {\bibfnamefont {Tomasz}\ \bibnamefont {Klimczuk}}, \ and\ \bibinfo {author} {\bibfnamefont {Bartlomiej}\ \bibnamefont {Wiendlocha}},\ }\bibfield  {title} {\enquote {\bibinfo {title} {Strong-coupling superconductivity of {SrIr}$_2$ and {SrRh}$_2$: Phonon engineering of metallic {Ir and Rh}},}\ }\href@noop {} {\bibfield  {journal} {\bibinfo  {journal} {Physical Review B}\ }\textbf {\bibinfo {volume} {104}},\ \bibinfo {pages} {054505} (\bibinfo {year} {2021})}\BibitemShut {NoStop}%
\bibitem [{\citenamefont {G{\'o}rnicka}\ \emph {et~al.}(2023)\citenamefont {G{\'o}rnicka}, \citenamefont {Winiarski}, \citenamefont {Walicka},\ and\ \citenamefont {Klimczuk}}]{gornicka2023superconductivity}%
  \BibitemOpen
  \bibfield  {author} {\bibinfo {author} {\bibfnamefont {Karolina}\ \bibnamefont {G{\'o}rnicka}}, \bibinfo {author} {\bibfnamefont {Micha{\l}~J}\ \bibnamefont {Winiarski}}, \bibinfo {author} {\bibfnamefont {Dorota~I}\ \bibnamefont {Walicka}}, \ and\ \bibinfo {author} {\bibfnamefont {Tomasz}\ \bibnamefont {Klimczuk}},\ }\bibfield  {title} {\enquote {\bibinfo {title} {Superconductivity in a breathing kagome metals {RO}s$_2$ ({R= Sc, Y, Lu})},}\ }\href@noop {} {\bibfield  {journal} {\bibinfo  {journal} {Scientific Reports}\ }\textbf {\bibinfo {volume} {13}},\ \bibinfo {pages} {16704} (\bibinfo {year} {2023})}\BibitemShut {NoStop}%
\bibitem [{\citenamefont {Wierzbowska}\ \emph {et~al.}(2005)\citenamefont {Wierzbowska}, \citenamefont {de~Gironcoli},\ and\ \citenamefont {Giannozzi}}]{wierzbowska}%
  \BibitemOpen
  \bibfield  {author} {\bibinfo {author} {\bibfnamefont {Malgorzata}\ \bibnamefont {Wierzbowska}}, \bibinfo {author} {\bibfnamefont {Stefano}\ \bibnamefont {de~Gironcoli}}, \ and\ \bibinfo {author} {\bibfnamefont {Paolo}\ \bibnamefont {Giannozzi}},\ }\bibfield  {title} {\enquote {\bibinfo {title} {{Origins of low-and high-pressure discontinuities of {T}$_c$ in niobium}},}\ }\href@noop {} {\bibfield  {journal} {\bibinfo  {journal} {arXiv preprint cond-mat/0504077}\ } (\bibinfo {year} {2005})}\BibitemShut {NoStop}%
\bibitem [{\citenamefont {Giustino}(2017)}]{gustino-rmp}%
  \BibitemOpen
  \bibfield  {author} {\bibinfo {author} {\bibfnamefont {Feliciano}\ \bibnamefont {Giustino}},\ }\bibfield  {title} {\enquote {\bibinfo {title} {Electron-phonon interactions from first principles},}\ }\href {\doibase 10.1103/RevModPhys.89.015003} {\bibfield  {journal} {\bibinfo  {journal} {Rev. Mod. Phys.}\ }\textbf {\bibinfo {volume} {89}},\ \bibinfo {pages} {015003} (\bibinfo {year} {2017})}\BibitemShut {NoStop}%
\bibitem [{\citenamefont {Allen}\ and\ \citenamefont {Dynes}(1975)}]{allen-dynes}%
  \BibitemOpen
  \bibfield  {author} {\bibinfo {author} {\bibfnamefont {P.~B.}\ \bibnamefont {Allen}}\ and\ \bibinfo {author} {\bibfnamefont {R.~C.}\ \bibnamefont {Dynes}},\ }\bibfield  {title} {\enquote {\bibinfo {title} {Transition temperature of strong-coupled superconductors reanalyzed},}\ }\href {\doibase 10.1103/PhysRevB.12.905} {\bibfield  {journal} {\bibinfo  {journal} {Phys. Rev. B}\ }\textbf {\bibinfo {volume} {12}},\ \bibinfo {pages} {905--922} (\bibinfo {year} {1975})}\BibitemShut {NoStop}%
\bibitem [{\citenamefont {Eliashberg}(1960)}]{eliashberg1960interactions}%
  \BibitemOpen
  \bibfield  {author} {\bibinfo {author} {\bibfnamefont {GM}~\bibnamefont {Eliashberg}},\ }\bibfield  {title} {\enquote {\bibinfo {title} {Interactions between electrons and lattice vibrations in a superconductor},}\ }\href@noop {} {\bibfield  {journal} {\bibinfo  {journal} {Sov. Phys. JETP}\ }\textbf {\bibinfo {volume} {11}},\ \bibinfo {pages} {696--702} (\bibinfo {year} {1960})}\BibitemShut {NoStop}%
\bibitem [{\citenamefont {Margine}\ and\ \citenamefont {Giustino}(2013)}]{PhysRevB.87.024505}%
  \BibitemOpen
  \bibfield  {author} {\bibinfo {author} {\bibfnamefont {E.~R.}\ \bibnamefont {Margine}}\ and\ \bibinfo {author} {\bibfnamefont {F.}~\bibnamefont {Giustino}},\ }\bibfield  {title} {\enquote {\bibinfo {title} {Anisotropic {M}igdal-{E}liashberg theory using wannier functions},}\ }\href {\doibase 10.1103/PhysRevB.87.024505} {\bibfield  {journal} {\bibinfo  {journal} {Phys. Rev. B}\ }\textbf {\bibinfo {volume} {87}},\ \bibinfo {pages} {024505} (\bibinfo {year} {2013})}\BibitemShut {NoStop}%
\end{thebibliography}%

\end{document}